\documentclass[longauth]{aa}

\usepackage{figsize}
\usepackage{graphicx}
\usepackage{txfonts}
\usepackage{xcolor}
\usepackage{url}
\usepackage{natbib}
\usepackage{multirow}
\usepackage{upgreek}
\usepackage{float}
\usepackage{placeins}
\bibpunct{(}{)}{;}{a}{}{,} 
\begin{document}

   \title{\texttt{CaRM}: Exploring the chromatic Rossiter-McLaughlin effect}
   \subtitle{The cases of HD 189733b and WASP-127b\thanks{Based on Guaranteed Time Observations collected at the European Southern Observatory under ESO programme 1102.C-0744 by the ESPRESSO Consortium.}}


   \author{         E. Cristo\inst{\ref{inst1},\ref{inst2}}
            \and   N. C. Santos\inst{\ref{inst1},\ref{inst2}}
            \and   O. Demangeon\inst{\ref{inst1},\ref{inst2}}
            \and J. H. C. Martins\inst{\ref{inst1},\ref{inst2}}
            \and P. Figueira\inst{\ref{inst1},\ref{inst6}}
            \and N. Casasayas-Barris\inst{\ref{inst3},\ref{inst4}}
            \and M. R.  Zapatero Osorio\inst{\ref{inst9}}
            \and F. Borsa\inst{\ref{inst5}}
            \and S. G. Sousa\inst{\ref{inst1}}
            \and M. Oshagh \inst{\ref{inst3},\ref{inst4}}
            \and G. Micela \inst{\ref{inst7}}
            \and H. M. Tabernero\inst{\ref{inst9}}
            \and J.V. Seidel \inst{\ref{inst19}}
            \and S. Cristiani\inst{\ref{inst10}}
            \and F. Pepe \inst{\ref{inst19}}
            \and R. Rebolo \inst{\ref{inst3},\ref{inst4},\ref{inst11}}
            \and V. Adibekyan \inst{\ref{inst1},\ref{inst2}}
            \and R. Allart \inst{\ref{inst19},\ref{inst20}}
            \and Y. Alibert \inst{\ref{inst12}}
            \and T. Azevedo Silva \inst{\ref{inst1},\ref{inst2}}
            \and V. Bourrier \inst{\ref{inst19}}
            \and A. Cabral \inst{\ref{inst14},\ref{inst16}}
            \and E. Esparza Borges\inst{\ref{inst3},\ref{inst4}}
            \and J. I. Gonz\'alez Hern\'andez \inst{\ref{inst3},\ref{inst4}}
            \and J. Lillo-Box \inst{\ref{inst9}}
            \and G. Lo Curto \inst{\ref{inst6}} 
            \and C. Lovis \inst{\ref{inst19}}
            \and A. Manescau \inst{\ref{inst17}}
            \and P. Di Marcantonio\inst{\ref{inst10}}
            \and C. J.A.P. Martins \inst{\ref{inst1},\ref{inst18}}
            \and A. S. Mascare{\~n}o \inst{\ref{inst3},\ref{inst4}}
            \and D. M{\'e}gevand \inst{\ref{inst19}}
            \and A. Mehner\inst{\ref{inst6}}
            \and N. J. Nunes \inst{\ref{inst14}}
            \and E. Palle \inst{\ref{inst4},\ref{inst21}}
            \and A. Sozzetti\inst{\ref{inst8}}
            \and S. Udry\inst{\ref{inst19}}
          }

    \institute{
          Instituto de Astrof\'isica e Ci\^encias do Espa\c{c}o, Universidade do Porto, CAUP, Rua das Estrelas, 4150-762 Porto, Portugal \label{inst1}
          \and
          Departamento de F\'isica e Astronomia, Faculdade de Ci\^encias, Universidade do Porto, Rua do Campo Alegre, 4169-007 Porto, Portugal \label{inst2}
          \and
                   European Southern Observatory, Alonso de C\'ordova 3107, Vitacura, Regi\'on Metropolitana, Chile\label{inst6}
        \and
          Instituto de Astrof\'{i}sica de Canarias (IAC), 38205 La Laguna, Tenerife, Spain \label{inst3}
          \and
          Universidad de La Laguna (ULL), Departamento de Astrof\'{i}sica, 38206 La Laguna, Tenerife, Spain\label{inst4}
         \and
        Centro de Astrobiolog\'\i a (CSIC-INTA), Crta. Ajalvir km 4, E-28850 Torrej\'on de Ardoz, Madrid, Spain \label{inst9}
        \and 
         INAF - Osservatorio Astronomico di Brera, Via Bianchi 46, 23807 Merate, Italy\label{inst5}
         \and
        INAF - Osservatorio Astronomico di Palermo, Piazza del Parlamento 1, 90134 Palermo, Italy\label{inst7}
        \and
        Département d’astronomie de l’Universit\'e de Gen\`eve, Chemin Pegasi 51, 1290 Versoix, Switzerland \label{inst19}
        \and 
        INAF - Osservatorio Astronomico di Trieste, via G. B. Tiepolo 11, I-34143 Trieste, Italy\label{inst10}
        \and 
                Consejo Superior de Investigaciones Cient\'{\i}cas, Spain\label{inst11}
        \and
        Department of Physics, and Institute for Research on Exoplanets, Universit\'e de Montr\'eal, Montr\'eal, H3T 1J4, Canada \label{inst20}
        \and
        Physics Institute, University of Bern, Sidlerstrasse 5, 3012 Bern, Switzerland\label{inst12}
        \and
        Instituto de Astrof\'isica e Ci\^encias do Espa\c{c}o, Faculdade de Ci\^encias da Universidade de Lisboa, Campo Grande, PT1749-016 Lisboa, Portugal \label{inst14}
        \and 
        Faculdade de Ci\^encias da Universidade de Lisboa (Departamento de F\'isica), Edif\'icio C8, 1749-016 Lisboa, Portugal \label{inst16}
        \and
        European Southern Observatory, Karl-Schwarzschild-Strasse 2, 85748  Garching b. M\"unchen, Germany \label{inst17}
        \and
        Centro de Astrof\'isica da Universidade do Porto, Rua das Estrelas, 4150-762 Porto, Portugal\label{inst18}
        \and
        Instituto de Astrofísica de Canarias (IAC), E-38200 La Laguna, Tenerife, Spain \label{inst21}
        \and
        INAF - Osservatorio Astrofisico di Torino, via Osservatorio 20, 10025 Pino Torinese, Italy \label{inst8}
}

   \date{Received XXXX xx, xxxx; accepted XXXX xx, xxxx}


  \abstract
  {}
   {In this paper we introduce \texttt{CaRM}, a semi-automatic code for the retrieval of broadband transmission spectra of transiting planets through the chromatic Rossiter-McLaughlin method. We applied it to HARPS and ESPRESSO observations of two exoplanets to retrieve the transmission spectrum and we analyze its fitting transmission models.}
   { We used the strong radius dependence of the Rossiter-McLaughlin (RM) effect amplitude, caused by planetary companions, to measure the apparent radius change caused by the exoplanet atmosphere. In order to retrieve the transmission spectrum, the radial velocities, which were computed over wavelength bins that encompass several spectral orders, were used to simultaneously fit the Keplerian motion and the RM effect. From this, the radius ratio was computed as a function of the wavelength, which allows one to retrieve the low-resolution broadband transmission spectrum of a given exoplanet. \texttt{CaRM} offers the possibility to use two Rossiter-McLaughlin models taken from \texttt{ARoME} and \texttt{PyAstronomy}, associated with a Keplerian function to fit radial velocities during transit observations automatically. Furthermore it offers the possibility to use some methods that could, in theory, mitigate the effect of perturbation in the radial velocities during transits.}
   {We applied \texttt{CaRM} to recover the transmission spectrum of HD 189733b and WASP-127b, with HARPS and ESPRESSO data, respectively. Our results for HD 189733b suggest that the blue part of the spectrum is dominated by Rayleigh scattering, which is compatible with former studies. The analysis of WASP-127b shows a flat transmission spectrum.}
   {The \texttt{CaRM} code allows one to retrieve the transmission spectrum of a given exoplanet using minimal user interaction. We demonstrate that it allows one to compute the low-resolution broadband transmission spectra of exoplanets observed using high-resolution spectrographs such as HARPS and ESPRESSO.}


   \maketitle
%

\section{Introduction}

Radial velocities (RVs) and photometry are the most efficient methods  used to detect and characterize exoplanetary systems\footnote{\url{www.exoplanet.eu}}. In particular, the first exoplanet detected orbiting a main-sequence star (using the RV method, \citealt{Mayor1995}) pushed forward the development of new state-of-art spectrographs \citep[e.g.,][]{Mayor2003, Pepe2021}. The improvement of instrumental and observational capabilities made it possible, not only to detect planets, but also to characterize them and, in particular, their atmospheres \citep[e.g.,][]{ Sing2009, Wyttenbach2015, Nikolov2018, Ehrenreich2020}. \par
During the transit of an exoplanet, the stellar radiation is filtered by the day-night terminator of the exoplanet's atmosphere. When crossing the stellar disk, the radiation is blocked by opacity sources such as atoms, molecules, and small dust particles, which mimic an increase in the radius of the planet \citep[e.g.,][]{2007ApJ...661L.191B,Seager2010}. By measuring the radius as a function of the wavelength, it is possible to recover the transmission spectrum of the planetary atmosphere  \citep{Seager2000}. \par
Transmission spectroscopy has now been broadly used to detect both narrow- and broadband features \citep[e.g.,][]{Kreidberg2014, Sing2015,2019A&A...627A.165H,2020A&A...641A.123H,Allart2020, Ehrenreich2020, CasasayasBarris2021,Borsa2021}. A notable example is the claimed discovery of the sodium doublet signature in the atmosphere of the exoplanet HD 209458b \citep{Charbonneau2002} with the NASA/ESA Hubble Space Telescope (HST), which opened a new era in the study of exoplanetary atmospheres \footnote{See the recent result from \cite{CasasayasBarris2021}.}. From the ground, the first molecular signature for the same element was detected by \cite{2008ApJ...673L..87R} from the optical transmission spectrum of the exoplanet HD 189733b. Furthermore, other molecules and atomic species such as H$_2$O, CO, CH$_4$, CO$_2$, Li, K, Fe, or TiO have been observed, both from space and the ground, in several exoplanets \citep{Wilkins2013, Wakeford2015, Sing2015, Tabernero2021}. \par
In broadband, Rayleigh scattering is responsible for the increase in opacity in the blue visible region of a spectrum, as a result of the interaction of small particles with light. The characteristic slope, which which originates in this mechanism, has been detected in several hot Jupiters \citep{Pont2008, Wakeford2013, Nikolov2018}. The Rayleigh and/or Mie scattering mask the presence of features from the deeper atmospheres, which include the broadened alkali K and Na species, particularly in the visible, but also in the near-infrared \citep[e.g.,][]{Wakeford2013, Wakeford2015, Sing2015}.\par
The Rossiter-McLaughlin (RM) effect \citep{Holt1893, Rossiter1924,McLaughlin1924} can be used as a tool to measure changes in the radius of an exoplanet with wavelength, the so-called chromatic Rossiter-McLaughlin \citep{2009A&A...499..615D}. This alternative way of retrieving a transmission spectrum was explored for the first time by \cite{Snellen2004}, where the author measured, from ground-based observations, the increased amplitude of the RM effect around Sodium D-lines when compared with the average amplitude for the exoplanet HD 209458b. Later, \cite{DiGLoria2015} applied the same technique to HD 189733b using the High Accuracy Radial velocity Planet Searcher (HARPS, \citealt{Mayor2003}) data and found the broadband signature of Rayleigh scattering, which has been confirmed by \cite{Oshagh2020}. \par
In this paper, we present \texttt{CaRM}, a novel implementation of the  chromatic Rossiter-McLaughlin effect (CRM). The CRM, although low resolution, has the advantage of capturing the broadband transmission spectrum of the exoplanet atmosphere. With high-resolution techniques, the “renormalization” of the photometric continuum wipes out any information it contains in broadband.\par
\texttt{CaRM} has already been used in \cite{Santos2020} for the Echelle SPectrograph for Rocky Exoplanets and Stable Spectroscopic Observations (ESPRESSO, \citealt{Pepe2010,Pepe2013,Pepe2021}) data of HD 209458b. In this work, \texttt{CaRM} is used to analyze HARPS and ESPRESSO data of HD 189733b and WASP-127b, respectively. In Sect. \ref{CaRM} we describe the principles behind the RM effect and how it can be used to measure the planet radius at different wavelengths. We also present our implementation of the CRM effect in detail. In Sect. \ref{mlbd}, we explore and assess the effect of considering different RM models and limb-darkening laws. Finally, we detail how we employed \texttt{CaRM} to retrieve the transmission spectrum of the exoplanets HD 189733b and WASP-127b (Sects. \ref{obs}, \ref{trans} and \ref{trans_wasp}). For the former, we modeled the transmission spectrum to, in addition, address the significance of the broadband feature that is observed.

\section{\texttt{CaRM}}\label{CaRM}
\subsection{The chromatic Rossiter-McLaughlin effect}\label{crm}

The RM effect is an anomaly in the RV curve of a star observed during a transit or eclipse. When the planet or star crosses the stellar disk, it covers regions with different stellar rotational RVs. This causes an unbalance of the integrated RVs for each element of the surface of the star, and a shift in its measured RV. The amplitude (and shape) of the RM effect contains information about the radius ratio between planet and star, the impact parameter, the sky-projected spin-orbit angle, and spin velocity of the star \citep{Triaud2018}:

\begin{equation}
A_{RM} \propto\frac{2}{3} \,D(\uplambda)\, v\,\sin \, i_\star \sqrt{1-b^2},
\end{equation}
where $v\,\sin\, i_\star$ is the equatorial velocity of the star projected over the line of sight, $D(\uplambda)=(R_{\rm p}(\uplambda)/R_{\rm \star})^2$, and $b$ the impact parameter which corresponds to the sky-projected distance between the center of the stellar disk and the position of the planetary disk at conjunction. Since the value of $D$ can be a function of wavelength ($\uplambda$), this equation shows that a measurement of the amplitude of the RM effect can be used to measure the planet radius variations as a function of the wavelength, that is to say to retrieve the transmission spectrum of the exoplanet.\par

\begin{table}
\caption{\texttt{CaRM} parameters. The first set corresponds to the orbital properties combined with the parameters that may help to model activity signals. The second constitutes the set of stellar properties, from which $u_i$, $v\,\sin \,i_\star$, $\beta_0$, and $\zeta_t$ are also employed for the RM modeling. }
\label{table:description}
\centering
\begin{tabular}{ll}
\hline
Parameter & Physical meaning \\
\hline \noalign{\smallskip}
$V_{\rm sys}\,$[km\,s$^{-1}$] & Systematic velocity of the system \\ \noalign{\smallskip}
$R_{\rm p}/R_{\rm *}$ & Radius ratio between planet and host star \\ \noalign{\smallskip}
$K\,$ [km\,s$^{-1}$]& Keplerian semi-amplitude \\ \noalign{\smallskip}
$a /R_\star$& Semi-major axis in units of stellar radius \\ \noalign{\smallskip}
$i_p\,$[$^\circ$]& Orbital inclination \\ \noalign{\smallskip}
$\lambda\,$[$^\circ$] & Spin-orbit angle \\ \noalign{\smallskip}
$\sigma_{\rm 0}\,$ [km\,s$^{-1}$]\tablefootmark{*}& Width of the best Gaussian fit to the CCFs \\ \noalign{\smallskip}
$\log(\sigma_{\rm W})$& Jitter logarithm amplitude \\ \noalign{\smallskip}
$\xi \,$ [km] &  Linear slope \\ \noalign{\smallskip}
$\log(a)$\tablefootmark{**} & Log-amplitude of the GP kernel \\ \noalign{\smallskip}
$\log(\tau)$ \tablefootmark{**} & Log-timescale of the GP kernel \\ \noalign{\smallskip}
$\Delta \phi_0$ & Shift in phase about the mid-transit \\ \noalign{\smallskip}
\hline
Stellar Properties & \\
\hline \noalign{\smallskip}
$v\,\sin i_\star \,$ [km\,s$^{-1}$] \tablefootmark{*}& Projected stellar rotation velocity \\ \noalign{\smallskip}
$u_i$ & Limb-darkening coefficients \\ \noalign{\smallskip}
$\beta_{\rm 0} \,$ [km\,s$^{-1}$] \tablefootmark{*} & Width of the nonrotating star \\ \noalign{\smallskip}
$\zeta_t \,$ [km\,s$^{-1}$]\tablefootmark{*}& Macro-turbulence amplitude \\ \noalign{\smallskip}
$(T_{\rm eff},\, \sigma_{\rm T_{\rm eff}})\,$ [K]& Effective temperature and uncertainty \\ \noalign{\smallskip}
$(\log(g),\, \sigma_{\rm \log(g)})$& Surface gravity and uncertainty \\ \noalign{\smallskip}
$(\rm[Fe/H],\, \sigma_{\rm [Fe/H]})\,$ [dex] & Metallicity and uncertainty \\ \noalign{\smallskip}
\hline
\end{tabular}
\tablefoot{\tablefoottext{*}{These parameters are replaced in the PyAstronomy RM model by the stellar angular velocity $\Omega$. Additionally it is necessary to provide the inclination of the stellar rotation axis $i_\star$. This model does not incorporate the macro-turbulence effect on the RM anomaly.}\\ \tablefoottext{**}$ $ Used for GPs.}
\end{table}

\begin{figure*}
  \centering
  \includegraphics[width=0.7\linewidth]{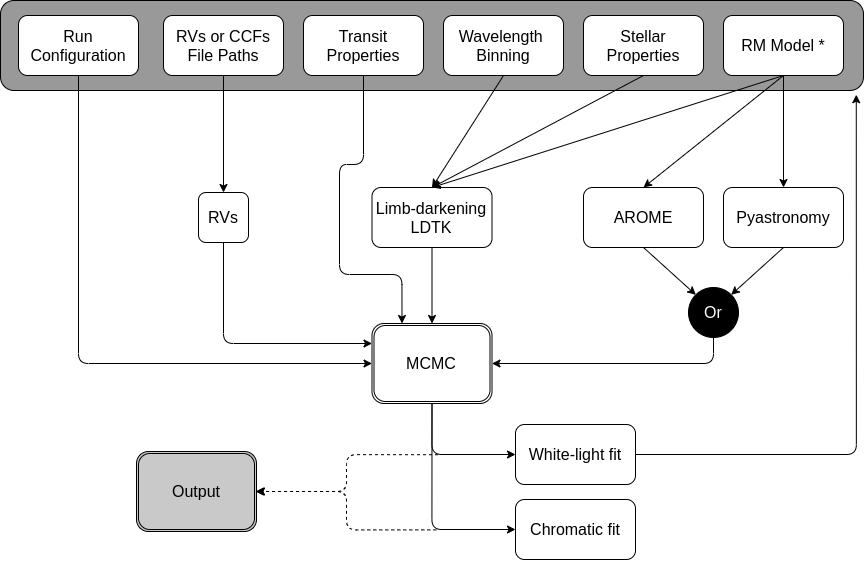} 
  
    \small\textsuperscript{*} Which may include or not GPs or the linear slope.
  \caption{Processes flow scheme of \texttt{CaRM}. The top gray area represents the user inputs and the bottom is for the last process execution in a run. The RVs can be directly read from text files or, alternatively, computed from the CCFs files for HARPS
and ESPRESSO. }
  \label{fig:scheme}
\end{figure*}
               
\texttt{CaRM}, for chromatic Rossiter-McLaughlin, is a code developed to measure amplitude variations of the RM anomaly as function of wavelength. For this, the user has to provide the RVs as a function of the wavelength. Then, the data in each wavelength bin, defined by the user, are fitted with an RM model. From the fit, the planet's radius is computed and, as a result, the broadband transmission spectrum of the exoplanet is returned.

\subsection{Data input}
The input data can be given in the form of standard CCF files, as delivered by their respective pipelines for ESPRESSO and HARPS, or text files (one for each bin). For ESPRESSO, the orders are duplicated thanks to the use of a pupil slicer delivering two simultaneous spectra (or two slices) for each spectral order \citep{Pepe2021}.\par
For the automated retrieval of the RVs for HARPS and ESPRESSO, the code takes as input the ranges of orders or slices, which define the bins, as well as the data paths and the DRS fits' files extensions for the CCFs. For each bin, it performs a flux weighted (by the inverse variance) sum of the corresponding CCFs. The flux uncertainties are computed with the standard error for the weighted mean for each point \footnote{$ \frac{1}{\sigma^2}= \sum_{i=1}^{n} \frac{1}{\sigma_{n}^{2}}$, for $n$ flux points with uncertainty $\sigma_n$.}. The RV corresponds to the mean of the Gaussian fit of the resulting CCFs associated to each wavelength bin. For ESPRESSO, the error is estimated by taking the gradient of the CCF flux (in the RVs' axis) and summing the inverse square of the former weighted by the CCF errors. For HARPS, there is no information about the uncertainty of the flux in each CCF point. In this case, we employed the \cite{Bouchy2001} method directly to each spectral order (which is read from the S2D files). Both methods are equivalent and are used to measure the contribution of the photon noise to the uncertainty in RV measurements. \par
For the text files, in the table format, the first line defines the wavelength range and the second provides the identification of each data column. The table itself is expected to contain the observation times (BJD), the RVs, and respective uncertainties (km\,s$^{-1}$). An example is provided together with the code\footnote{\url{https://https://github.com/EduardoCristo/CaRM}}.\par

\subsection{The models}
The RVs of a star hosting planets are composed of the joint velocity of the system, the Keplerian motion, and, during the transit, the RM effect. In \texttt{CaRM}, the Keplerian component is modeled by a sinusoidal function (i.e., in this version only a circular orbit is considered) with a semi-amplitude  $K$,  an orbital period $P$, and a shift of the mid-transit time, in terms of orbital phase, $\Delta \phi_{\rm 0}$. A constant velocity, $V_{\rm sys}$, is added to account for the systemic velocity. For the "RM" component, we employed two distinct models: \texttt{ARoME}  \citep{Boue2013} and \cite{Ohta2005} (implemented in \texttt{PyAstronomy}, \citealt{Czesla2019}). \par
The measurement of the Doppler shift of spectra can be made using, for example, the iodine cell technique \citep{Butler1996} or the CCF technique that uses the cross-correlation between the observed spectrum and a mask that indicates the location and strength of the known stellar lines  \citep{Baranne1996}. \texttt{ARoME} is a code developed to accommodate these observation techniques. The version we included in \texttt{CaRM} is an adaptation with \texttt{Cython} of the original library \footnote{Available on: \url{http://www.astro.up.pt/resources/arome/}}. As inputs, the code uses the planet to star radius ratio ($R_{\rm p}/R_{\rm \star}$), the mid-transit mean-anomaly that is derived from the phase assuming an argument of periastron of $\pi / 2$, the semi-major axis of the orbit ($a/R_\star$), the orbital inclination ($i_p$), the projected angle between the spin-axis  and the orbital plane ($\lambda$), the projected rotational velocity, and the stellar macro-turbulence parameter $\zeta_{\rm t}$. This last parameter combined with the stellar rotation velocity is added to the width of the Gaussian fit to the stellar surface CCF behind the planet ($\beta_{\rm 0}$) plus the width that results from the Gaussian fit to the out-of-transit CCFs ($\sigma_{\rm 0}$). \par
The model available in the \texttt{PyAstronomy} package \citep{pya} uses a geometrical approach to describe the RM anomaly based on the description of \cite{Ohta2005} for RVs derived from template matching. The projected velocity toward the observer is modeled as the sum of the projected proper motion and the rotational contribution of the surface. The model uses the equatorial stellar angular velocity ($\Omega$), which we computed using the rotational velocity of the star ($V_{\rm rot}$) with a fixed stellar radius, and the inclination of the stellar rotation axis ($i_\star$).\par
Both RM models require one to provide a set of limb-darkening coefficients. For \texttt{ARoME}, it accepts both quadratic or nonlinear limb-darkening coefficients as input. In contrast, \texttt{PyAstronomy} only works with a linear law. The impact of using different limb-darkening laws is explored in Sect.\ref{mlbd}. We use the limb-darkening toolkit \texttt{LDTk}\footnote{\url{https://github.com/hpparvi/ldtk}} \citep{Parviainen2015}, which is a package that automates the computation of the limb-darkening profiles using specific intensity spectra generated from the \texttt{PHOENIX} spectral library \citep[][]{Husser2013}. This code allows the fit of limb-darkening profiles in specific, user provided, wavelength bins. To retrieve the coefficients, \texttt{LDTk} uses, in addition to the wavelength interval, the effective temperature of the star $T_{\rm eff}$, the logarithm of the stellar surface gravity $\log(g)$, and the stellar metallicity [Fe/H]. We adopted an approach similar to the photometric transmission spectroscopy, where the limb-darkening profile is calculated integrating over the full range of the wavelength bin. A word of caution is needed, though, since the RM effect is a consequence of the shape variation of the stellar lines in a given wavelength range. Wavelength trends on the strength of individual spectral lines would change the limb-darkening profiles, which can potentially mimic chromatic variations. \par
\texttt{CaRM} also offers the possibility to model the activity-induced RV signal using two approaches: adding a linear phase-dependent slope to the data and adding a GP model. The former takes into account long-term trends due to stellar rotation and the effect of stellar activity in the out-of-transit slope, while the latter can be used to try to model in-transit variability (e.g., occulted spots or granulation).\par
For the linear slope, if the transit window is provided, the code can identify the in-transit and out-of-transit measurements. This is fitted simultaneously with the RM curve.\par
For the GP analysis, \texttt{CaRM} calls the {\tt celerite}  $RealTerm$ kernel \citep{ForemanMackey2017}. This requires the definition of the log of amplitude and timescale, $log(a)$ and $log(\tau)$, of the residual RV signal. The GP kernel takes the following functional form:
\begin{equation}
    k(\phi)=a e^{-\phi \tau},
\end{equation}
where $\phi$ is the orbital phase.
\subsection{Implementation}\par
\texttt{CaRM} is developed in a blend of \texttt{Python3} and \texttt{Cython3} with common standard packages such as \texttt{Numpy} and \texttt{Scipy}. The RM models are chosen by the user employing an input file. This file consists of a set of instructions to fit both the white light (i.e., using the full passband of the instruments) and the chromatic data. There we define the stellar and orbital properties combined with the selection of the RM model. As input, we can select the preferred limb-darkening law and provide stellar parameters such as the radius, the effective temperature, log-surface gravity, and metallicity (Table \ref{table:description} and the documentation available on \texttt{CaRM}'s page). Furthermore, in the same file, we provide the inputs and the initial estimates of the model parameters, or if each parameter must be fitted or not with a given prior. The range of the priors and the possibility of some parameters to be independently fit of each night can be enabled. The code assumes that if the prior is not declared, the corresponding parameters are fixed with the value provided by the user input values.\par
A call of the code starts to build the “results” file in which the parameters provided by the user are stored (see Fig. \ref{fig:scheme}). The times of observation for each epoch are converted into orbital phases using the input values for the period and the mid-transit time. The passbands and stellar parameters are used to compute the limb-darkening coefficients, for the specified law, with \texttt{LDTk}. For the combined white light and chromatic analysis, the first data set that is run is for the white-light RVs. After this fitting, taking advantage of the higher S/N, it is possible to use the computed parameters to constrain the chromatic run (assuming they are wavelength independent). The best-fit model values can be obtained, in an automated way and  by user definition, as the posterior maximum probability  or median  solutions. 
The prior definition, the chains (which are picked from a randomly drawn distribution from the priors), and log likelihood are provided to the Markov chain Monte Carlo (MCMC) implementation \texttt{emcee} \citep{ForemanMackey2013}. Each evaluation of the model is computationally expensive; however, with the parallelization potentialities, the sampling is accelerated, reducing the overall time of computation. \par
The MCMC runs in two distinct phases. The burn-in is used to sample the posterior distributions and narrow the range of values of each parameter.
After this, the final position of the walkers is saved and used as the initial position for the production phase. The fit of the chromatic data has a similar process. The major distinction is the use of the updated parameters, fixing the constant parameters from the white-light RV fitting. After the code runs for all the passbands, the information is stored for analysis.

 \section{The limb-darkening impact}\label{mlbd}
Limb darkening is a radial effect produced by the observation of atmospheric stellar layers at different temperatures. From the perspective of the observer, the effective optical depth decreases with increasing radius due to lower gas density and a shorter line-of-sight distance through the star.
\begin{figure}[h!]
  \centering
  \includegraphics[width=\linewidth]{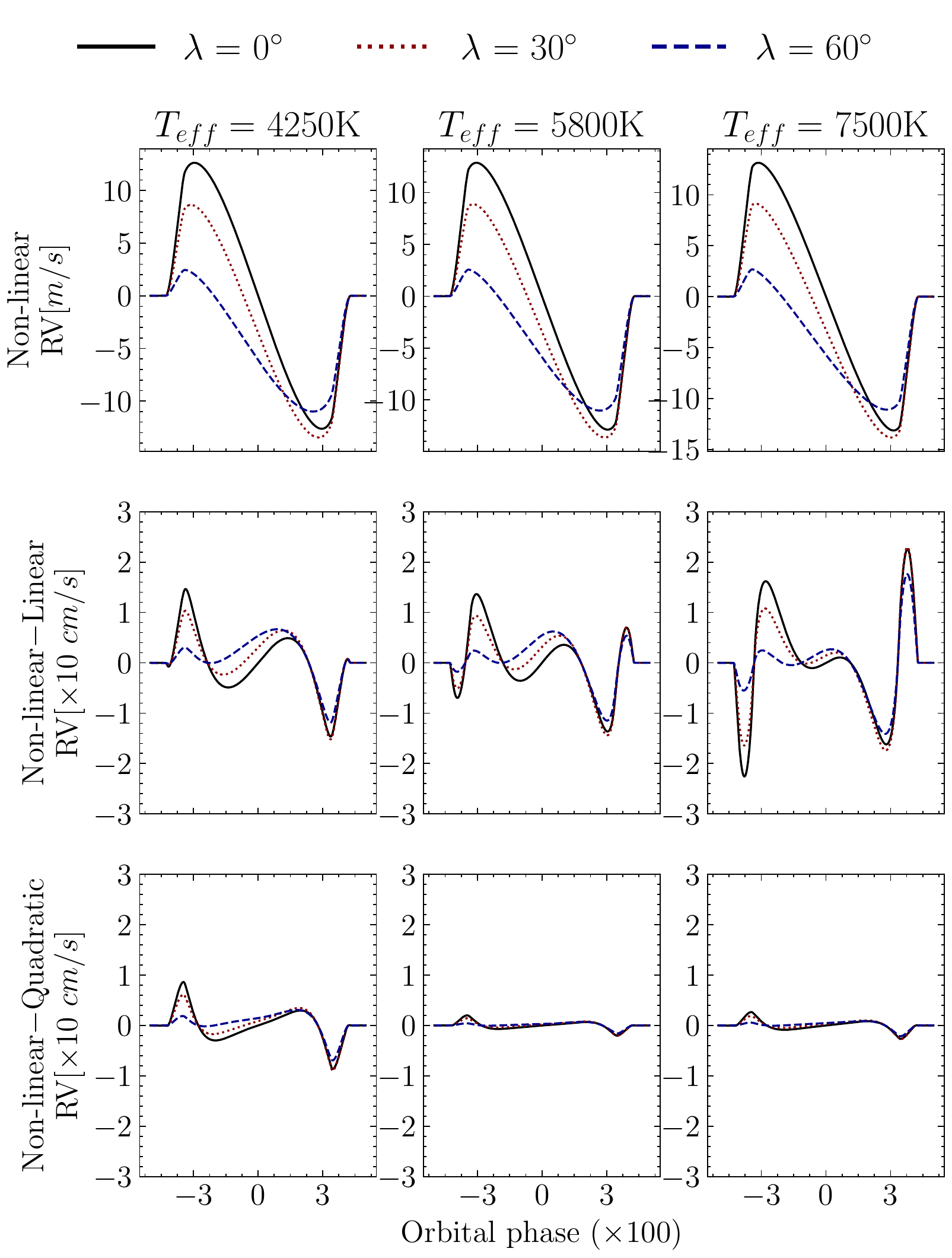}
  \caption{ Stellar temperature impact, in RVs, in the limb-darkening laws. Top row: RM curve (computed with \texttt{ARoME}), with increasing spin-orbit misalignment, for three  temperatures in the temperature range of the FGK stars. Middle and bottom row: Residuals' variation (in tenths of cm/s) between a linear and a quadratic law. The RV scale is constant to emphasize the variation of the residuals' amplitude with stellar temperature.}
  \label{fig2:crmdif}
\end{figure}
\begin{figure*}[h!]
  \centering
  \includegraphics[width=\linewidth]{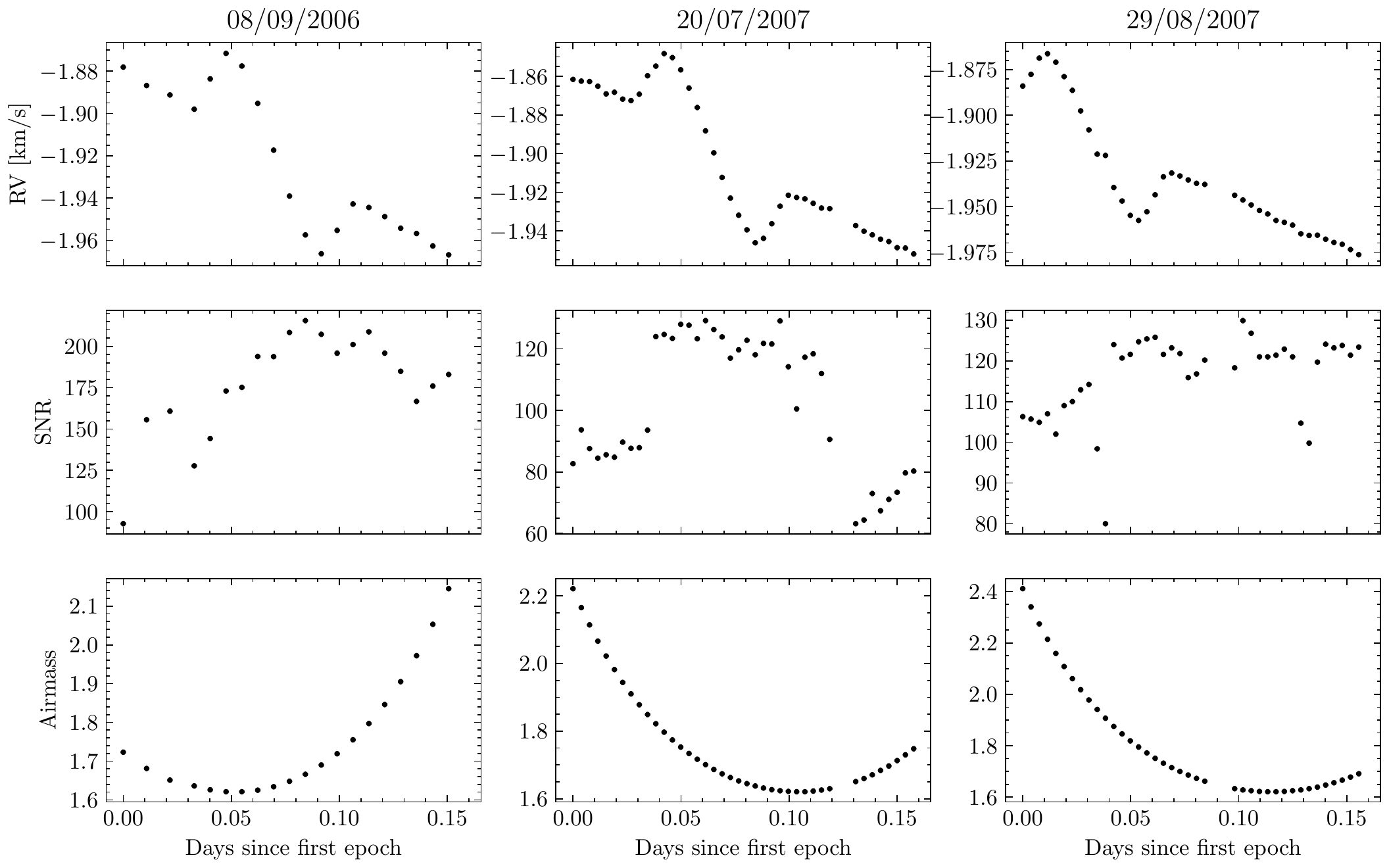}
  \caption{RVs of HD 189733b retrieved from the CCF header for the white light. The error bars are not visible since they are smaller than the dimension of the markers. In the same column, the associated S/N near $580$ nm and the airmass is a function of the number of days since the first epoch. The average signal-to-noise is $178$, $102,$ and $116,$ respectively.}
  \label{fig:hd189733bcombined}
\end{figure*}
\cite{Espinoza2016} studied the impact of the different limb-darkening laws (sect. \ref{ldlaws}) in transit light curves. They conclude that, comparing the limb darkening recovered for a number of Kepler targets, there are biases introduced that are particularly relevant for noncentral transits. This can affect the determination of the stellar inclination, semi-major axis ($a/R_{\rm \star}$), and the planet radius ($R_{\rm p}/R_{\rm \star}$). The proposed solution to partly solve these biases is to let the limb-darkening coefficients free with high polynomial order limb-darkening laws. \par
To verify if the choice of the limb-darkening law can produce false transmission spectroscopy features, we assessed the impact in RVs, assuming that the intrinsic law that models the stellar limb-darkening profile of the simulated system is nonlinear. For this, we fit the limb-darkening laws (whose profiles are compared in Fig. \ref{fig1:profdif}) with the \texttt{LDTk} package for a simulated typical system with a hot Jupiter (parameters in Table \ref{table:tab1}). The impact of the laws in the RM  was computed with the average temperature range (i.e., from minimum to maximum) of the FGK stars. We then compared the Rossiter-McLaughlin anomalies, computed with \texttt{ARoME},  to quantify the differences in RVs between linear and quadratic laws with the underlying nonlinear law (Fig. \ref{fig2:crmdif}).\par

The RVs' comparisons show the same global behavior, with the quadratic law presenting the lowest residuals overall. In addition, there is an increasing trend in the amplitude of the linear residuals, whilst there is a slight decrease for the quadratic law. ESPRESSO, for example, has achieved a precision of $< 30 $  cm\,s$^{-1}$  \citep{2020A&A...639A..77S} on multiple observations, despite the precision being able to be lower for single measurements (depending, e.g., on the magnitude or spectral type). For the simulated hot-Jupiter system, the peak difference ($< 25 $  cm\,s$^{-1}$) is below the precision of most state-of-the-art spectrographs. As such, we would not be able to measure differences that originate from the choice of the limb-darkening law and which could produce some chromatic variability.

\section{Observations and data reduction}\label{obs}
To apply \texttt{CaRM} to the observations of HD 189733b and WASP-127b, we computed the RVs using the CCF method. The CCFs from each observation were computed by the HARPS and ESPRESSO data reduction software packages (DRS version 3.4 and 2.2.8, respectively), and they are provided per  order and per slice (ESPRESSO).
\begin{figure*}
  \centering
  \includegraphics[width=\linewidth]{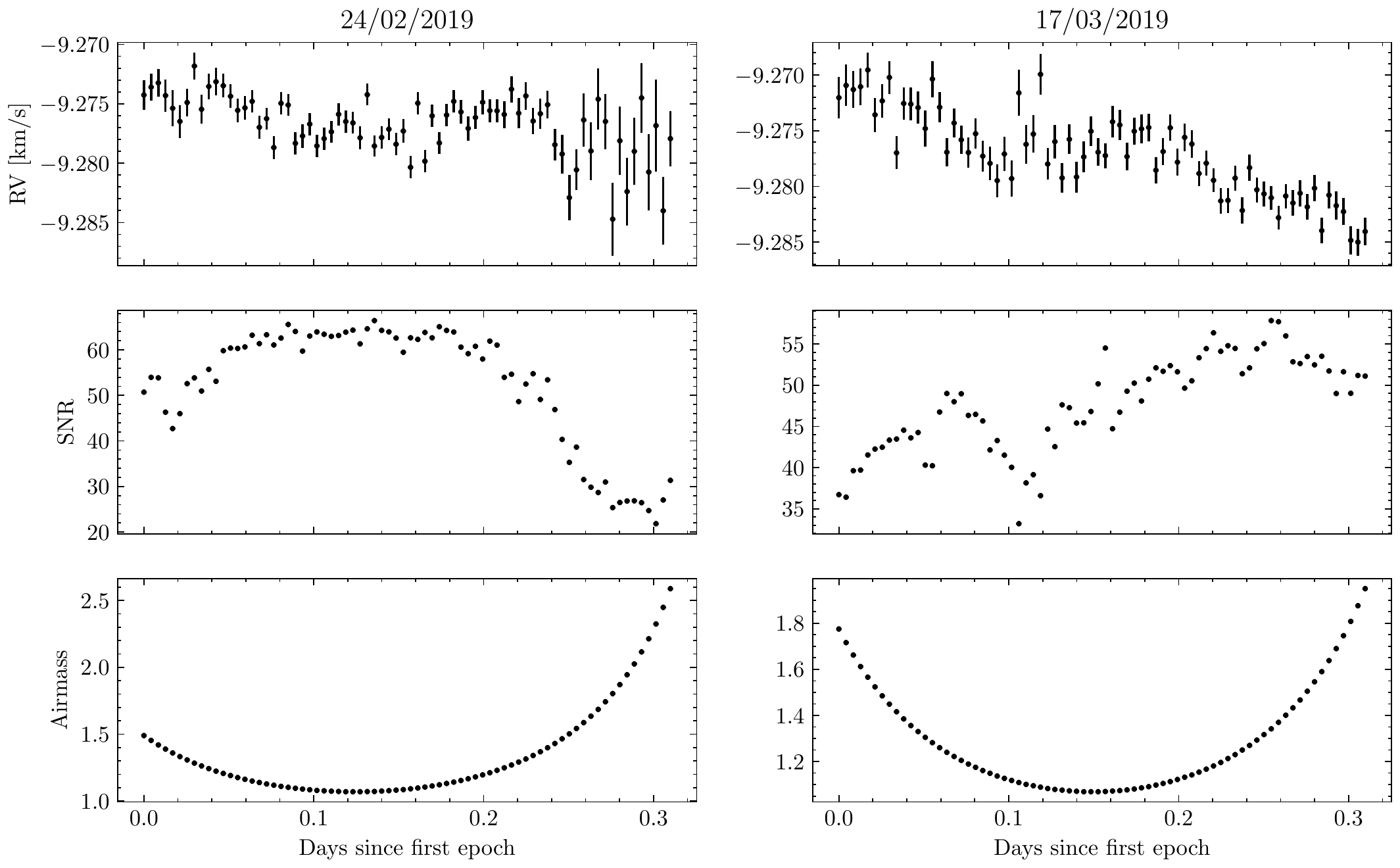}
  \caption{RVs retrieved from the CCF header for the white light and respective error bars for each night for WASP-127b. In the same column, the associated S/N near $580$ nm and the airmass is a function of the number of days since the first epoch.}
  \label{fig:wasp127data}
\end{figure*}
\subsection{HD 189733b}
The HD 189733b data consists of three sets of observations taken on 8 September 2006, as well as 20 July and 29 August 2007 with HARPS (Fig. \ref{fig:hd189733bcombined}). This corresponds to the same data analyzed in \cite{DiGLoria2015}. HARPS is a fiber-fed spectrograph installed on the $3.6$ m telescope Cassegrain in La Silla. It covers a spectral range between $378\,$nm and $691\,$nm, distributed over $71$ echelle orders. The observations were acquired under the programs 072.C-0488(E), 079.C-0828(A) (PI:Mayor), and 079.C-0127(A) (PI: Lecavelier des Étangs). The exposure time was set to $900$s for the first night and $300$s for the following one. The detector was set for the fast readout and low gain mode, which offers a lower overhead between exposures ($23$s when compared with $180$s for the high gain mode) at the expense of increased readout noise. \par
 
The CCFs were produced cross-correlating with a K0 spectral template binary mask, closely matching the spectral type of the host star, HD 189733 \citep[K2 V from][]{Gray2003}. This choice is is different from that that adopted by \cite{DiGLoria2015}, which used the reduction available in the ESO archive. That reduction used a different mask, G2, for the last two nights.\par

\subsection{WASP-127b}
The observations of WASP-127b were acquired on the 24 February and 17 March 2020 under the GTO program ID 1102.C-0744 (PI:
F. Pepe). In total, there are 148 RV measurements with $300$s of integration. The spectrograph was set to singleHR21 mode with a spectroscopic HR (which corresponds to the highest prescribed RV precision of the instrument), one UT, slow reading, and $2 \times 1$ binning. The RVs were extracted using a G8 mask as provided by the ESPRESSO DRS. This same data were analyzed by \cite{Allart2020} to study the atmosphere of WASP-127b in high resolution.\par
The inspection of the data shows hints of a low amplitude RM effect. The after transit RVs of the first night reveal an increased dispersion and average error when compared with the rest of the data. This is accompanied by a decrease in the signal-to-noise (S/N) and it is caused by an increase in the airmass (see Fig. \ref{fig:wasp127data}).\par

\section{The transmission spectrum of HD 189733b}\label{trans}
\subsection{The white-light RM fit}\label{hd189sub1}

As presented in \cite{Santos2020} for HD 209458b, the first step to retrieve the transmission spectrum is to fit the white-light data with the RM model. This is performed to fix wavelength-independent parameters using the higher S/N data. We used the \texttt{ARoME} model and adopted, as a reference, most parameters from \cite{Triaud2009}, which is the same source from the non model-specific parameters used by \cite{DiGLoria2015} (see Table \ref{table:tablepars}). We also used the updated stellar parameters, with the exception of the stellar radius, derived in \cite{Sousa2008} and listed in SWEET-CAT \footnote{\url{https://www.astro.up.pt/resources/sweet-cat/}} \citep{Santos2013} which compiles sets of stellar atmospheric parameters derived, when possible, with a homogeneous methodology for planet host stars.\par
\begin{table}
\centering
\caption{Set of priors for the white-light fit with \texttt{ARoME} for HD 189733b. In the table $\mathcal{G}$ represents a normal distribution where the first value corresponds to the mean and the second is for the standard deviation. Also $\mathcal{U}$ represents the uniform distribution with the respective lower and upper boundaries. The chromatic posteriors are not represented in the table as they change from bin to bin; the corner plots are in section \ref{appendix}. }
\label{hd189priors}
\begin{tabular}{lll}
Parameter & Prior \\
\hline
$V_{\rm sys}\,$ [km\,s$^{-1}$] & $\mathcal{U}(-2.2,-1.6)$\\
$K \,$[km\,s$^{-1}$] & $\mathcal{G}(0.20196,0.1)$\\
$R_{\rm p} /R_{\rm \star} $ & $\mathcal{G}(0.1581,0.01581)$\\
$v\,sin\,i_{\rm \star} \,$[km\,s$^{-1}$] & $\mathcal{G}(3.05, 0.1)$\\
$\sigma_{\rm W} \,$[m\,s$^{-1}$] & $\mathcal{G}(0, 25)$\\
\hline
\hline
\end{tabular}
\end{table}
The initial value for the mean systematic velocity of the star $V_{\rm sys}$ ($-1.9$ km\,s$^{-1}$) was estimated by observation of the data, and it is only a rough estimate. In Fig. \ref{fig:hd189733bcombined} we see that, taking the average position of the mid-transit as a reference, this value is quite variable. This variation is expected between observations as a consequence of the nightly zero point, which can have an origin in the different calibrations performed at the beginning of each night for different observations, or as a result of stellar variability\footnote{We note that this star is chromospherically active \citep[][]{Bouchy2005, Boisse2009}.}. We chose to give a comprehensive uniform prior to account for these variations.\par
\begin{table}
    \centering
        \caption{Posterior distribution of the white-light fitted parameters for the preferred solution of HD 189733b. The values correspond to the median of the distribution and the uncertainties correspond to a $68 \%$ confidence interval. The subscript numbers represent the posterior median of the values that were fitted independently for each observation. }
    \label{table:hd189_posterior}
    \begin{tabular}{lll}
        Parameter & Value & Uncertainty\\
        \hline
        $V_{\rm sys,\,0}$ [km\,s$^{-1}$]  &$-1.91934$ &$\pm{0.00019}$\\
        $V_{\rm sys,\,1}$ [km\,s$^{-1}$]  &$-1.89739$ &$\pm{0.00016}$\\
        $V_{\rm sys,\,2}$ [km\,s$^{-1}$]  &$-1.91172$ &$\pm{0.00034}$\\
        $R_{\rm p}/R_{\rm *}$           &$0.18250$  &${+0.0017}/{-0.0016}$\\
        $K_{\rm 0}$ [km\,s$^{-1}$]        &$0.2069$   &${+0.0015}/{-0.0014}$\\
        $K_{\rm 1}$ [km\,s$^{-1}$]        &$0.2107$   &$\pm{0.0013}$\\
        $K_{\rm 2}$ [km\,s$^{-1}$]        &$0.1838$   &$\pm{0.0019}$\\
        $v\,sin\,i_\star$ [km\,s$^{-1}$]  &$3.253$    &$\pm{0.05}$\\
        $\sigma_{\rm W,\,0}$ [m\,s$^{-1}$]&$1.06$     &$\pm{0.14}$\\
        $\sigma_{\rm W,\,1}$ [m\,s$^{-1}$]&$1.32$     &${+0.13}/{-0.14}$\\
        $\sigma_{\rm W,\,2}$ [m\,s$^{-1}$]&$2.08$     &$\pm{0.14}$\\
        \hline
        \hline
    \end{tabular}
\end{table}
The  spin-orbit angle prior was fixed to the \cite{Triaud2009} estimate using the classic RM approach, $-0.85^{+0.28}_{-0.32}$ $^{\circ}$ (using also the same dataset). Other estimates from literature, using different data sets or techniques,  such as $-1.4 \pm 1.1 ^{\circ}$ from \cite{Winn2006},  $-0.35\pm 0.3 ^{\circ}$ and $-0.67 \pm 0.3 ^{\circ}$ (independently estimated from two transits) from \cite{CollierCameron2010} employing the line profile tomography technique, or $-0.4 \pm 0.2 ^{\circ}$ from \citet{Cegla2016} who used the reloaded RM technique, are compatible with the one adopted and they suggest an aligned orbit. 
The projected rotational velocity of the star ($v\,sin\,i_\star$) was first estimated by \cite{Bouchy2005} using the ELODIE spectrograph, from the following line broadening method: $3.5 \pm 1.0$ km\,s$^{-1}$. \cite{Cegla2016} recovered the rotational profile of HD 189733 and excluded the rigid body rotation with a high degree of confidence. They discuss that considering the rigid body rotation in these cases biases the velocities' ($v\,sin\,i_\star$) computation toward a lower value, and that it can produce wave-like residuals such as the ones observed by \cite{Triaud2009}. We chose to fit the velocity with a Gaussian prior centered on the value from \cite{Triaud2009} after the correction (minimizing the RM residuals), as the RM models used by \texttt{CaRM} do not include stellar differential rotation.\par
The value adopted for $\sigma_0$, that is, the width of the best Gaussian fit, results from the averaged measurement of the out-of-transit CCF widths obtained from a Gaussian fit. We estimate a similar value as \citep{DiGLoria2015}, but it is important to note that there is a variability of $0.03$ km\,s$^{-1}$ between the average out-of-transit of the night with the highest and lowest value. This should not  greatly impact the fit; nevertheless, the RM amplitude does not have a strong dependence on this parameter and the variation represents a small percentage of the value we used \citep[][]{Santos2020}. The width of the nonrotating star $\beta_0$ depends on measurements of the CCF width of a sample of stars and incorporates the instrumental profile. We used the calibration done for HARPS data, similar to the one presented in \citet[][]{2002A&A...392..215S}. \par
\begin{figure}[t!]
  \centering
  \includegraphics[width=\linewidth]{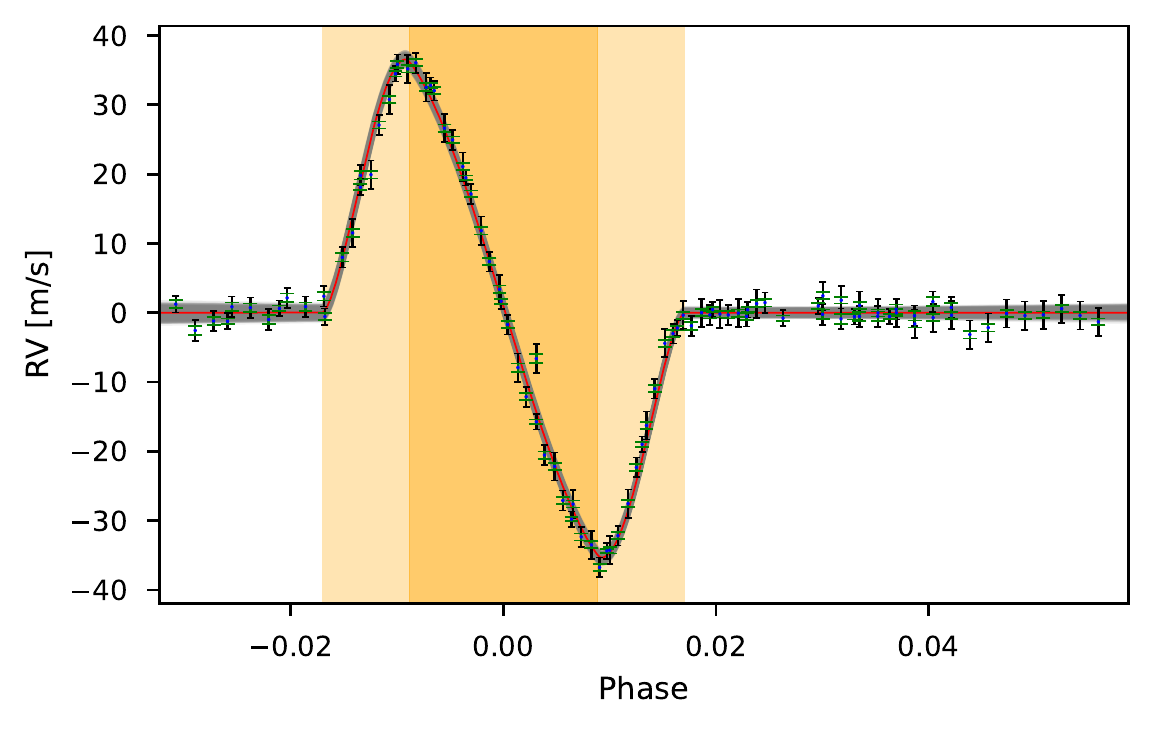}
  \includegraphics[width=\linewidth]{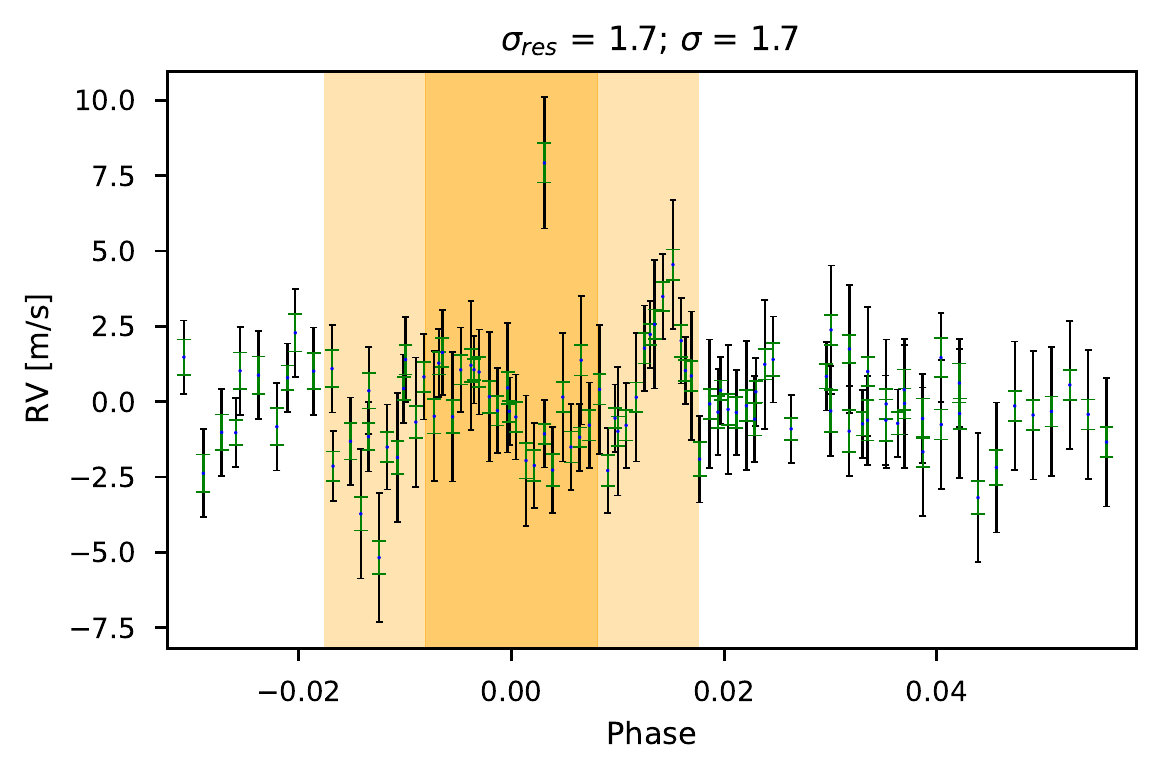}
  \caption{HD 189733 RM fit. Top: Best fit model (solid red) of the combined observations of HD 189733b with HARPS. The gray area corresponds to a random sample of the posterior about the highest likelihood. The data points present two error bars, the green corresponding to the original RVs' uncertainties computation from the CCF method and the black after quadratically adding the white-noise level fitted to the data. Bottom: Residuals after subtraction of the data from the best-fit. The lighter yellow represents the ingress and egress and the darker the region where the planet is fully in front of the stellar disk. At the top of the residual's plot, $\sigma_{\rm res}$ represents the average value of the RVs' residual dispersion, $\sigma_W$ is the estimate of the jitter velocity, and $\sigma$ is the average uncertainty in the residual RVs.}
  \label{fig:hd189jres}
\end{figure}
The white light was set to run with $50$ chains with an initial burn-in phase with $1500$ steps and $3000$ for the production, using as an initial guess the parameters from Table \ref{table:tablepars} and the priors from Table \ref{hd189priors}.
\begin{figure*}[h!]
  \centering
  \includegraphics[width=\linewidth]{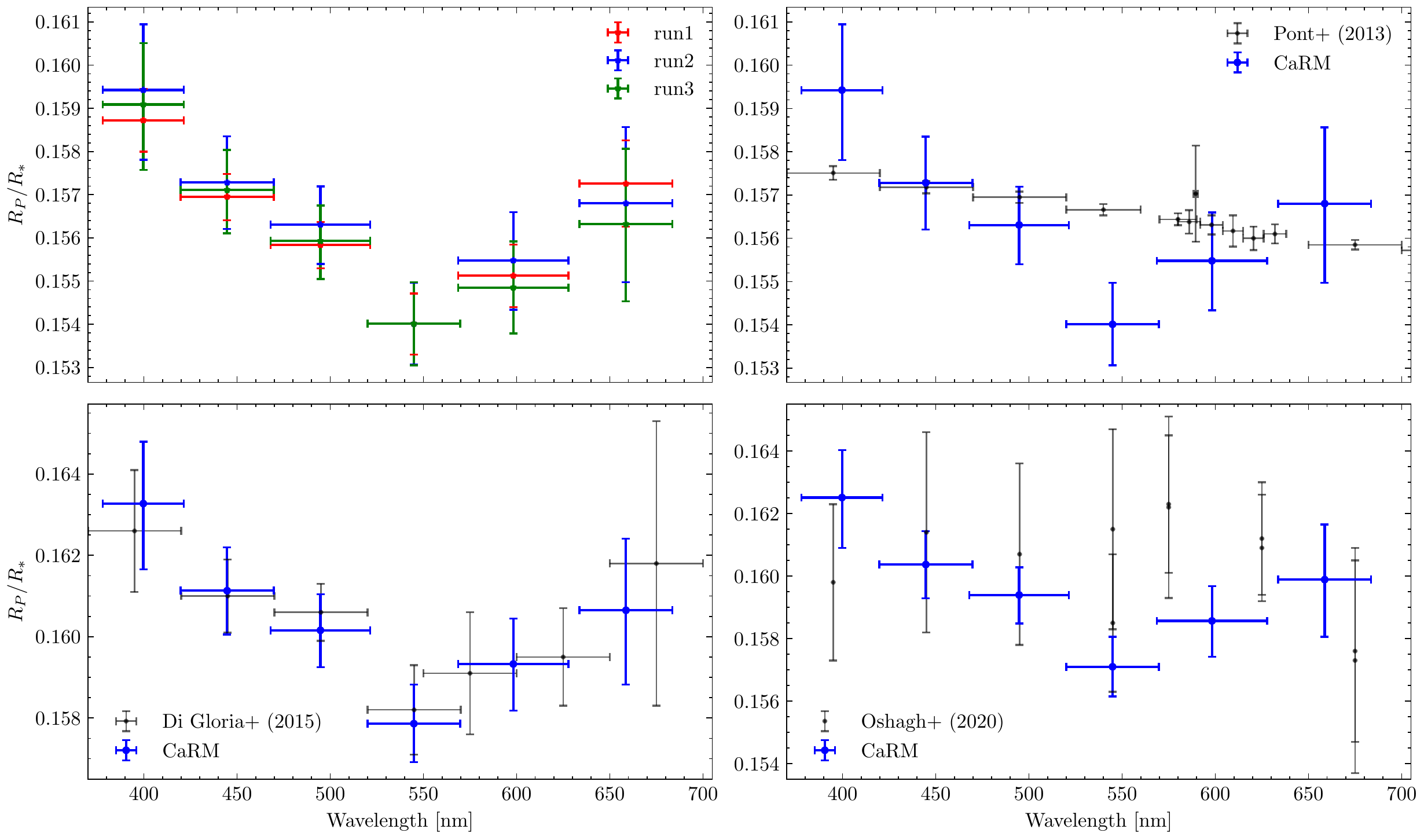}
  \caption{HD 189733b transmission spectra. Top left: Comparison of the transmission spectra retrieved with \texttt{CaRM}, with \texttt{ARoME}, for different setups (see text). The bins centered near $545$nm are aligned for better comparison. After, a comparison between our result and \cite{Pont2013} observed with Hubble (STIS), \cite{DiGLoria2015} with HARPS, and \cite{Oshagh2020} which combines HARPS and CARMENES data. Shifts where applied to compare with the various literature sources.}
  \label{fig:hdothers}
\end{figure*}
The individual fits of the three observations and residuals after subtraction of the best fit model are presented in Fig. \ref{fig:hd189nights}. By inspection, there is some structure in the residuals of each night that turns out to be more evident when the residual RVs are combined (Fig. \ref{fig:hd189jres}). The HARPS data of HD 189733b have been extensively analyzed in the literature \citep[e.g.,][]{Triaud2009, DiGLoria2015, Cegla2016}, and similar residual structures have been found. Some hypotheses that were proposed to explain this are differential rotation, spot crossing, or convective blueshift (CB). \par
The RM models we include assume a rigid body rotation for the star, which is equivalent to fix the stellar surface rotational velocity. It is known that stars possess differential rotation, in particular the Sun, which is a function of the latitude. \cite{Cegla2016} addressed this issue with the reloaded RM, which does not assume any specific line profile (and then velocity profile) to the occulted areas by the planet. They find that with differential rotation in the model, the wave-like features vanish. This star has also been reported to be active \citep{Bouchy2005} and multiple events of spot crossing have been detected in photometric transits \citep{Pont2013}. In the context of the RM, similar spot-crossing events (phase aligned) seem unlikely with observations that are taken far apart in time. Furthermore, the effect would have to be present during the complete duration of the transit, as the residuals have the same time span. Finally, CB \citep[e.g.,][]{2011ApJ...733...30S}, which is caused by the disk-integrated differences in brightness from the hotter upward-flowing material and the downward-flowing one, is a hypothesis that also needs to be considered. In \texttt{ARoME}, the effect of the CB is said to be negligible, as it only adds a constant offset to the RVs and does not modify the RM signal. Despite this, \cite{2012ApJ...757...18A} shows that although symmetric, it contributes to distortions in amplitude of the RM signal as a result of distortions of the line profiles from the center to the stellar limb \citep{Dravins2021}.

The median of the posterior distribution and uncertainties of the fitted parameters are in Table \ref{table:hd189_posterior}. The $K$ parameter,
 with a contribution from the Keplerian amplitude and variability, differs the most for the third night of observations. During this night, the beginning of the transit was not observed and roughly half of the data points correspond to out-of-transit measurements (Fig. \ref{fig:hd189nights}). It can be seen in the residuals, in Fig. \ref{fig:hd189jres}, that structures with an amplitude up to $\sim 2.5$ m\,s$^{-1}$  do not correspond to in-transit events. The higher level of activity also affects the jitter amplitude, which is the highest from the three nights. The median value of $\beta_0$ is in line with the value reported by \cite{DiGLoria2015}. For the projected rotational velocity, the median value of the posterior seems to tend to a higher value, standing between the value reported by \cite{Triaud2009} and \cite{Bouchy2005}.

\subsection{Transmission spectrum}
The hot Jupiter HD 189733b has been extensively studied since its discovery \citep{Bouchy2005}. The bluer side of the optical transmission spectrum of HD 189733b was reported to present a high radius-wavelength slope \citep{Pont2008, Sing2008, DiGLoria2015} which is not often seen in the transmission spectra of exoplanets of this type, a possible signature of Rayleigh scattering. Furthermore, there is no evidence for the detection of wings of the sodium doublet, while their cores are detected \citep[e.g.,][]{Wyttenbach2015, 2020EPSC...14..273S}. With the low amplitude water feature, the presence of an extended atmosphere dominated by submicron haze particles seems to dominate the spectra from visible to infrared wavelengths \citep{Sing2008, Sing2009}. \par
\begin{figure*}[ht]
\centering
  \includegraphics[width=0.49\linewidth]{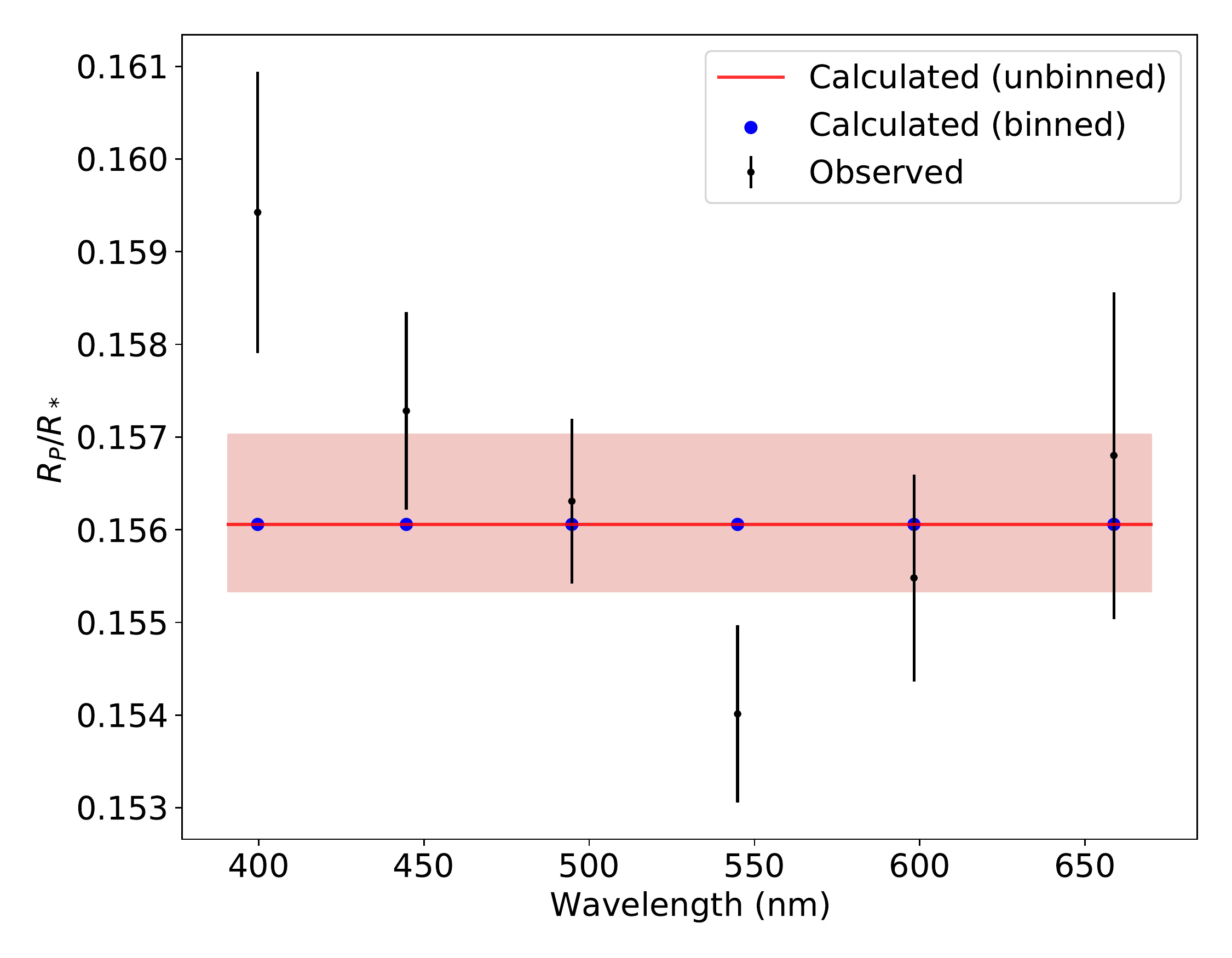}
  \includegraphics[width=0.49\linewidth]{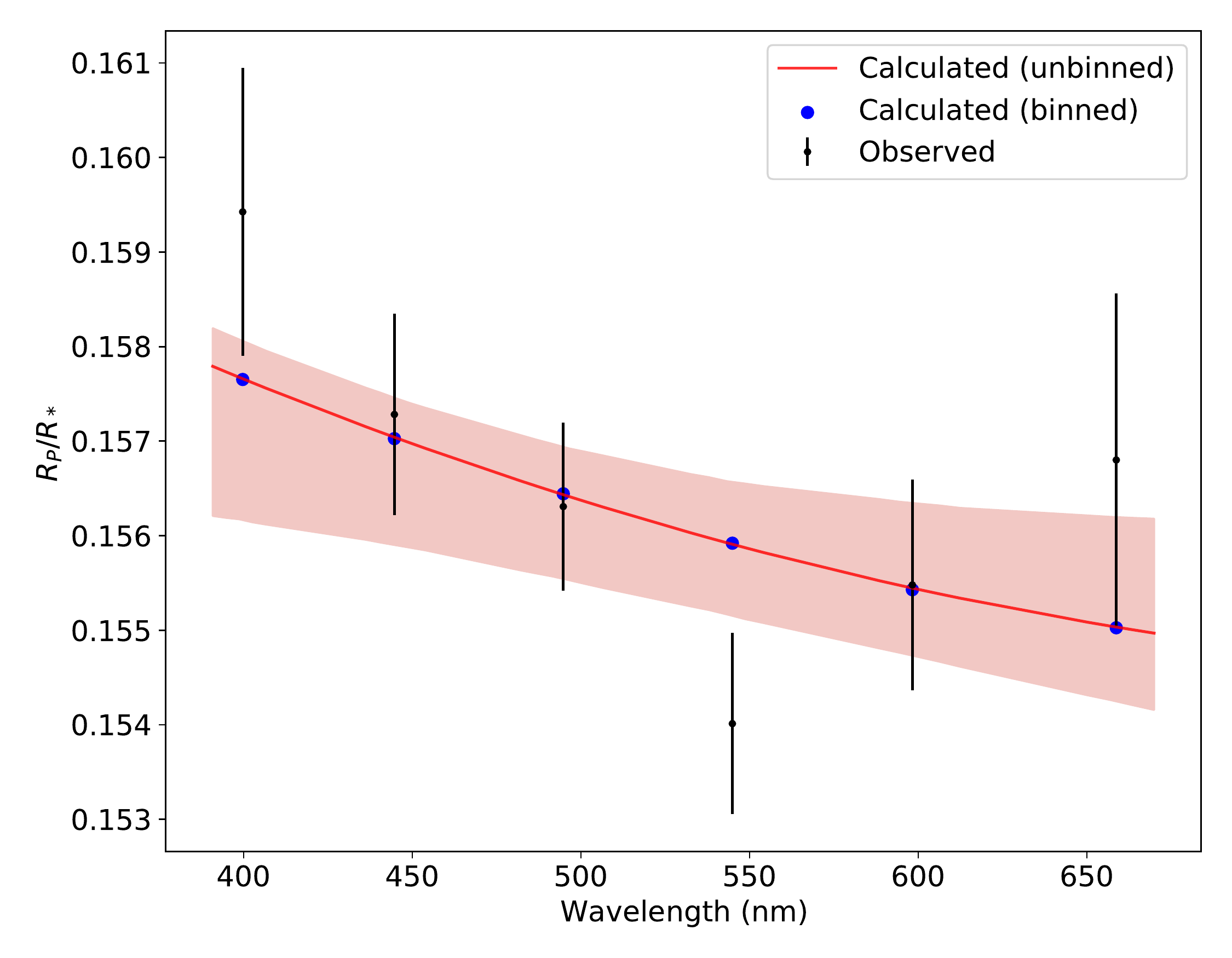}
\caption{Atmospheric models' fit of the preferred transmission spectrum with a flat profile (left) and adding Rayleigh scattering (right).}
\label{fig:hdtransmodels}
\end{figure*}
The chromatic Rossiter-McLaughlin method was applied twice to this target in the literature. The work of \cite{DiGLoria2015} found, with the same HARPS data analyzed here, the same strong wavelength radius slope consistent with the observations of \cite{Pont2008}. The value of the slope from HARPS data using the RVs obtained through the CCF method is larger when compared with the HST observations. Later, the study of \cite{Oshagh2020} shows that it is possible to obtain a slope closer to the one present on the  transmission spectrum derived from HST data, using RVs obtained via template matching \citep{Butler1996, AngladaEscude2012}. In the same paper, the hypothesis is raised that the standard templates from the HARPS DRS can contain blended lines that contribute to changes in the retrieved RVs, contaminating the transmission spectrum. Alternatively, \cite{Oshagh2014} and \cite{Boldt2020} show that stellar activity, such as spots of faculae, can mimic broadband features such as the slopes from Rayleigh scattering mechanisms. In particular, the activity seems to affect the RVs' observations producing variable slopes. Nevertheless, the probability of the activity giving an origin to a coherent (in time) strong feature is low, in particular with data from several transits.\par
To retrieve the transmission spectrum of HD 189733b, we used the standard RVs, computed with the CCF method with the default DRS masks (K0). 
As described in section \ref{CaRM}, we can use two methods that help to mitigate the impact of stellar activity in the retrieved transmission spectrum. In \cite{Santos2020}, the slope $K$ was set as a free parameter (for each night). \cite{DiGLoria2015} and \cite{Oshagh2020} fitted the out-of-transit points of each night separately and removed a linear trend before fitting the RM profile. In this context, $K$ must be interpreted not purely as the Keplerian amplitude, but a combination of it with the time-dependent linear slope.
\begin{table}
\centering
\caption{Set of priors for the chromatics fit with \texttt{ARoME} for HD 189733b.}
\label{table:hd189_prior_chrom}
\begin{tabular}{lll}
Parameter & Prior \\
\hline
$V_{\rm sys}\,$ [km\,s$^{-1}$] & $\mathcal{U}(-2.2,-1.6)$\\
$K \,$[km\,s$^{-1}$] & $\mathcal{G}(0.20196,0.1)$\\
$R_{\rm p} /R_{\rm \star} $ & $\mathcal{G}(0.1581,0.01581)$\\
$\sigma_{\rm W} \,$[m\,s$^{-1}$] & $\mathcal{G}(0, 50)$\\
\hline
\hline
\end{tabular}
\end{table}
The in-transit observation can also be affected by activity. Potential occultations of stellar spots or faculae are observed in the residuals, with a profile similar to the RM anomaly. This is seen in Fig. \ref{fig:hd189jres} and discussed in section \ref{hd189sub1}. \cite{DiGLoria2015} tried to mitigate the effect of these residuals by subtracting the residuals that result from the white-light fit from each chromatic bin. With \texttt{CaRM,} we chose not to perform this as it does not account for the chromatic variations of the phenomenon that gives origin to it. Furthermore, the contribution of the convective blueshift is differential along the stellar disk, causing center-to-limb variations \citep{Cegla2016, Reiners2016}. The line profile is deeper and thinner at the disk center and broader and shallower at the limb. The subtraction of the  white-light line profiles from the chromatics creates residuals of variable width  and depth \citep{Dravins2021}.\par
We performed three runs with \texttt{CaRM} to assess the impact on the transmission spectrum of different setups:
\begin{itemize}
    \item Gaussian processes with a linear slope $\xi$ (run1);
    \item fit the Keplerian semi-amplitude $K$ (run2);
    \item fit the linear slope (run3).
\end{itemize}
The parameters in these runs were fitted for each observation independently, and the transmission spectra are compared in Fig. \ref{fig:hdothers}. We preferred run2  (chromatic fit priors in Table \ref{table:hd189_prior_chrom}) since it minimizes the number of parameters in the model and, at the same time, can account for the chromatic variations during transit. The first runs present problems when fitting the out-of-transit slope in particular for the third night, as there is no available data before ingress. Despite this,  the comparison between retrievals show a good agreement at the $1\sigma$ level (Fig. \ref{fig:hdothers} and Table \ref{table:hd189_radius_table} for the preferred solution).\par
To check the validity of fixing parameters from the white light to the chromatic runs, in particular the ones associated with the \texttt{ARoME} model, we decided to measure the average out-of-transit CCF width as a function of the wavelength. We found a positive slope from blue to red wavelengths, with an average value which corresponds to the white-light $\sigma_0$ (Table \ref{table:tablepars}). To check if this variation affected the transmission spectrum, we provided as input to the code the measured $\sigma_0$ instead of fixing it to the white-light value. Since $\beta_0$ is correlated with $\sigma_0$, these variations also affect its value. We then quadratically added the difference between the bin CCF width and the white-light value. There were no significant changes, within the error bars, on the final transmission spectrum. In addition to this, we also checked if the transmission spectrum could be the result of wavelength-dependent BIS variations. For this, we subtracted the white-light BIS from the value for each bin and compared it with the general shape of the transmission spectrum. The BIS value corresponds to a few percent of the CCF widths; however, we did not find a correlation with the radius variations.\par

\subsubsection{Transmission spectrum modeling with \texttt{PLATON} }
To analyze the preferred transmission spectrum, we used \texttt{PLATON} \citep{Zhang2019, Zhang2020}. This tool is a \texttt{Python} package that computes transmission spectra and allows one to fit the transmission models to retrieve atmospheric characteristics using observational data \citep{Zhang2020}.\par
As a template to run the retrieval, we used the publicly available code example \footnote{\url{https://github.com/ideasrule/platon}}. We considered an isothermal atmospheric profile, which constitutes an adequate approximation \citep[e.g.,][]{2008A&A...481L..83L,Heng2017} for the atmospheric layers probed by transmission spectroscopy in the visible, with equilibrium chemistry. The stellar and planetary radius were set with a normal distribution, with the standard deviations and mean set to the values provided in Table \ref{table:tablepars}, taking a broader radius prior ($3\sigma$) to account for shifts in the average radius of the transmission spectrum. We used the value reported by \cite{Triaud2009} for the planet mass ($\mathcal{G}(1.138, 0.025)$) and the posterior values reported in \cite{Zhang2020} for the other parameters of HD 189733b. We used the priors  $\mathcal{G}(1089, 120)$ for the temperature of the atmosphere, $\mathcal{G}(1.08, 0.23)$ for the logarithm of the metallicity, and $\mathcal{G}(0.66, 0.09)$ for the C/O ratio. We increased the range of the uniform prior for the logarithm of the scattering factor and maintained the error multiple, $\mathcal{U}(-2, 6)$ and $\mathcal{U}(0.5, 5)$, respectively. The stellar temperature is fixed in the run, with the value also reported in Table \ref{table:tablepars}.\par
We tested the retrieval comparing the evidence of a flat model (no radius variation with wavelength) with one containing Rayleigh scattering, fixing the scattering slope parameter to $4$ (Fig. \ref{fig:hdtransmodels}), and a combination of Rayleigh scattering and Na. The Bayesian evidence was measured for each scenario using the built-in nested sampler to fit the transmission models. The data suggest the model with Rayleigh scattering is preferred over the flat model (with a logarithmic Bayesian evidence of $\ln(z)=32.4 \pm 0.1$ versus $\ln(z)=31.7 \pm 0.1$). The model considering sodium does not present any major difference with the one considering only Rayleigh, and the retrieval outputs an intermediate logarithmic evidence ($\ln(z)=32.2 \pm 0.1$). The transmission spectrum we obtained presents a good agreement when compared with \cite{DiGLoria2015}. Nevertheless, the evidence is not sufficient (with a Bayes factor of $\Delta \ln(z)=0.7 \pm 0.1$), due to the low resolution, to confirm the presence of Rayleigh scattering following the \cite{Kass95bayesfactors} criteria (which requires $\Delta \ln(z) > 3$).

\section{The transmission spectrum of WASP-127b}\label{trans_wasp}

\subsection{The white-light fit of WASP-127b}
\begin{figure}[h!]
  \centering
  \includegraphics[width=\linewidth]{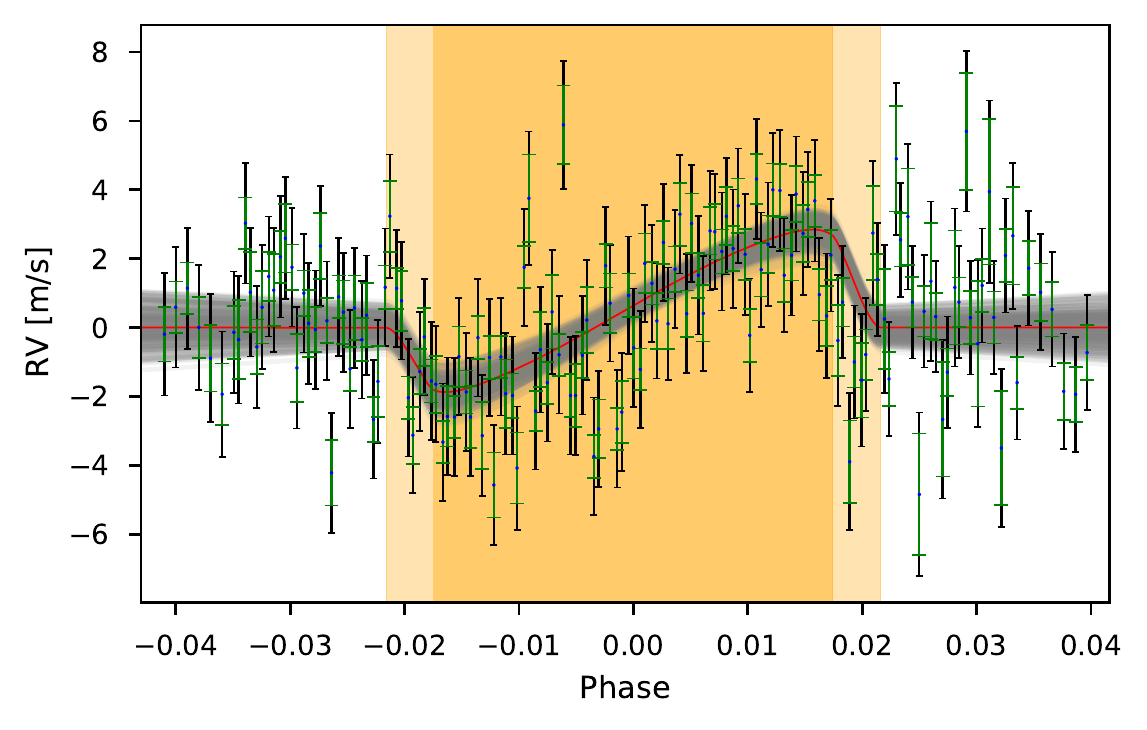}
  \includegraphics[width=\linewidth]{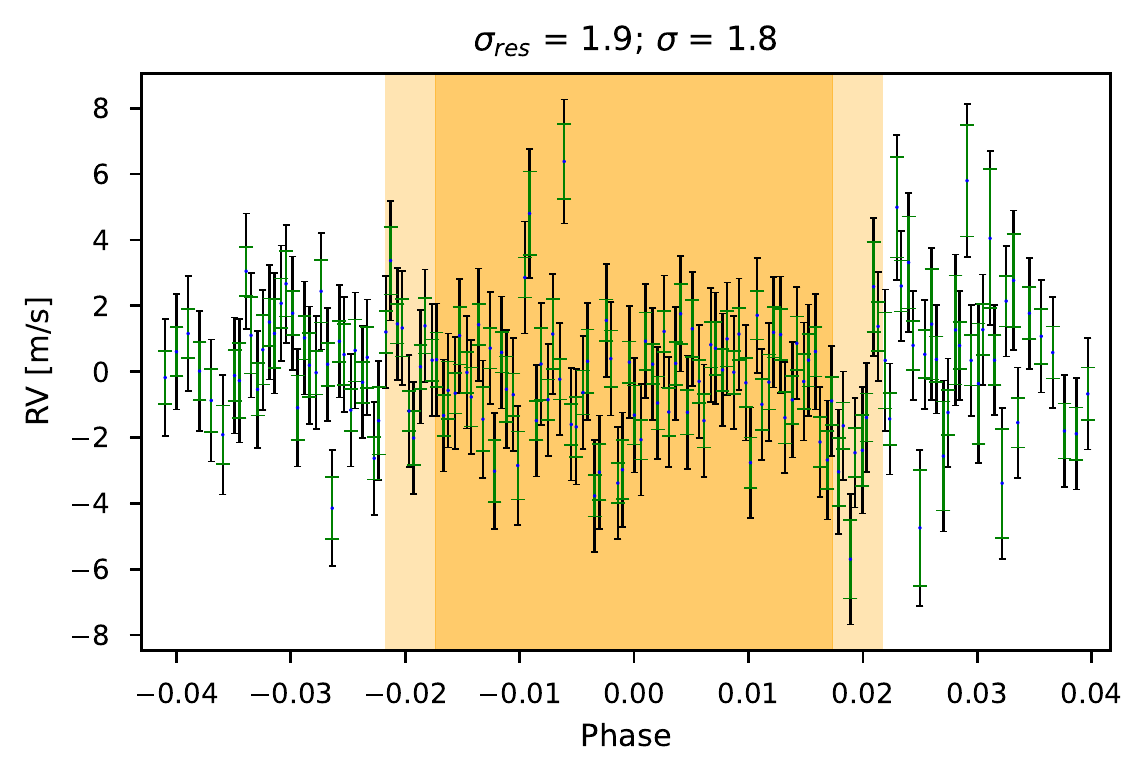}
  \caption{Combined data fit and residuals. The elements in the plots have the same meaning as in Fig. \ref{fig:hd189jres}.}
  \label{fig:wasp127jres}
\end{figure}

Similarly to the analysis of HD 189733b, we started fitting the combined observation in the full wavelength range of ESPRESSO  with the \texttt{ARoME} RM model. The $V_{\rm sys}$ encompasses a comprehensive interval around the observed average. The $R_{\rm p} /R_{\rm \star} $ prior is a Gaussian distribution centered in the value provided by \cite{Seidel2020} and the standard deviation corresponds to the $1\sigma$ uncertainty. The spin-orbit misalignment angle, in contrast with the previous study, has no published value using the classic RM analysis. It has been discussed in \citet{Cegla2016} that the reloaded RM analysis tends to give distinct values for the angle, which is attributed to center-to-limb convective variations. For a slow rotating star, such as WASP-127, we found that the residuals are dominated by these variations (producing amplitudes $\sim$ cm s$^{-1}$ to $\sim$ 1 m s$^{-1}$) and uncertainties in the obliquity from $10 ^{\circ} \sim 20 ^{\circ}$. It was chosen, as a result, to let the spin-orbit angle vary using a normal distribution prior with the mean from \cite{Allart2020} and a standard deviation of $20 ^{\circ}$ to account for the differences in the methods (priors summarized in Table \ref{tab:wasp127priors}). The MCMC run parameters were set in the same way as for HD 189733b for all bins. The combined data with the best fit model, as well as the residuals after subtraction from one to the other, are provided in Fig. \ref{fig:wasp127jres}.\par

\begin{table}
\centering
\caption{Set of priors for the white-light fit with \texttt{ARoME} for WASP-127b. }
\begin{tabular}{lll}
Parameter & Prior \\
\hline
$V_{\rm sys} \,$[km\,s$^{-1}$] & $\mathcal{U}(-9.775,-8.775)$\\
$R_{\rm p} /R_{\rm \star} $ & $\mathcal{G}(0.10103,0.010103)$\\
$\lambda \, [^\circ] $& $\mathcal{G}(-128.4,20)$\\
$v\,\sin\,i_{\star} \,$ [km\,s$^{-1}$] & $\mathcal{G}(0.53, 0.07)$\\
$\xi \,$ [km] & $\mathcal{G}(0, 0.1)$\\
$\sigma_{\rm W} \,$ [m\,s$^{-1}$] & $\mathcal{G}(0, 25)$\\
\hline
\hline
\end{tabular}
\label{tab:wasp127priors}
\end{table}
\begin{figure*}[ht!]
  \centering
  \includegraphics[width=\linewidth]{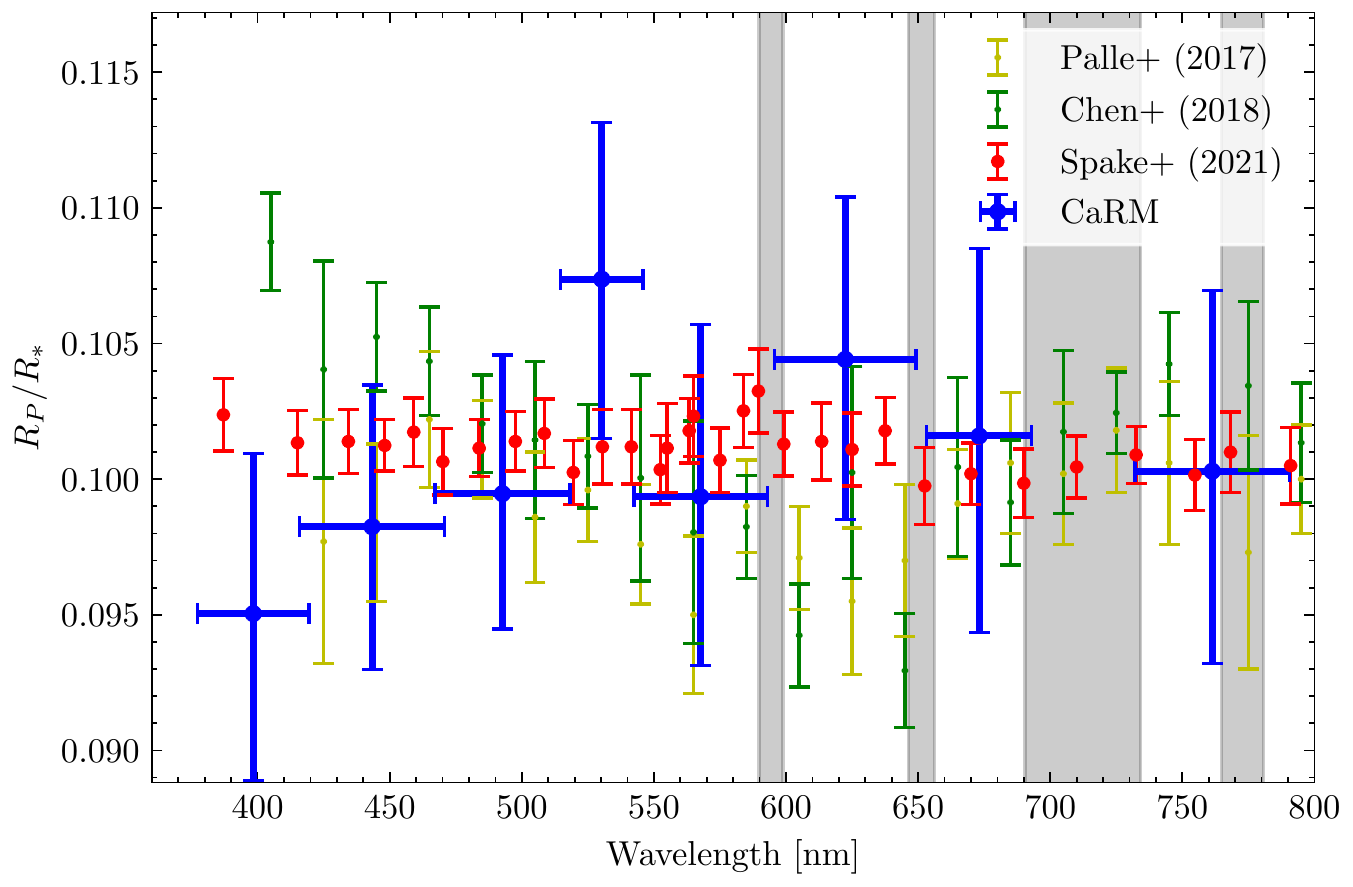}
  \caption{WASP-127b transmission spectrum retrieved with \texttt{CaRM}, with \texttt{ARoME}, compared with other sources in the literature (\citealt{Palle2017} with NOT/ALFOSC; \citealt{Chen2018} with NOT/ALFOSC and GTC/OSIRIS; and \citealt{Spake2021} with HST/STIS and TESS). The gray bars represent the wavelength intervals where the CCFs were masked to mitigate atmospheric contamination. The horizontal bars, in our retrieval, correspond to the binning in wavelength.}
  \label{fig:wasptransmission}
\end{figure*}
The posterior distribution of the fitted parameters for the white light is found in Table \ref{table:wasp127b_posterior}. The median systemic velocities for the observed nights present very similar values. The host star is, in fact, very similar to the Sun, but, with $9.7 \pm 1.0$ Gyr and a larger radius \citep{Allart2020}, it is at the end of its main-sequence phase and evolving toward the branch of the subgiants. Despite this, the star seems to present a low variability from these observations taken $21$ days apart. The spin-orbit angle value, $-150.6^{+8.1}_{-8.7}$ $^{\circ}$, has a difference of approximately $20 ^\circ$ when compared with the value from \cite{Allart2020} (Table \ref{table:tablepars}). This is at the top of the estimated interval by \cite{Cegla2016} for the  difference between the values obtained from the classical RM when compared to the reloaded RM. Despite this, the angle solutions are compatible at the $68\%$ level.

\begin{table}
    \centering
        \caption{Posterior distribution of the white-light fitted parameters for WASP-127b. The elements of the table have the same meaning as before.}
    \label{table:wasp127b_posterior}
    \begin{tabular}{lll}
        Parameter & Value & Uncertainty \\
        \hline
        $V_{\rm sys,\,0} \,$ [km\,s$^{-1}$]   &$-9.27656$   & $\pm {0.0020}$\\
        $V_{\rm sys,\,1}\,$ [km\,s$^{-1}$]    &$-9.27694$   & $\pm {0.0018}$\\
        $R_{\rm p}/R_{\rm *}$               &$0.0974$    & ${+0.0046}/{-0.0045}$\\
        $\lambda \,[^\circ]$                &$-150.6$    & ${+8.1}/{-8.7}$\\
        $\xi_{\rm \,0} \,$ [km\,s$^{-1}$]     &$0.028$      &  $\pm {0.011}$\\
        $\xi_{\rm \,1} \,$ [km\,s$^{-1}$]     &$-0.016$     &$\pm {0.010}$\\
        $\sigma_{\rm W,\,0} \,$ [m\,s$^{-1}$] &$1.58$       & ${+0.17}/{-0.16}$\\
        $\sigma_{\rm W,\,1} \,$ [m\,s$^{-1}$] &$1.47$       & ${+0.14}/{-0.15}$\\
        \hline
        \hline
    \end{tabular}
\end{table}

\subsection{The transmission spectrum of WASP-127b}
WASP-127b is an exoplanet that orbits a bright star in a short orbital period (Table \ref{table:tablepars}). In terms of proximity to the star and radius, it can be classified as a hot Jupiter; however, in terms of mass, it resembles more a member of the Neptune population. With 0.18 Jupiter masses \citep{Lam2017} and just over a Jupiter radius, it is one of the lowest density planets known. Its high insolation and low mass places it in the so-called Neptune desert \citep{Etangs2006,Mazeh2016}.
\cite{Lam2017} estimate a large-scale height for this exoplanet atmosphere ($\sim 2350$ km), which makes it a good target for transmission spectroscopy. Despite this, the rotational velocity of the star is low. The amplitude of the Rossiter-Mclaughlin effect is just $\sim 3$ \,m\,s$^{-1}$, which is an order of magnitude lower than the amplitude of the RM detected on HD 189733b.\par
The first complete study of WASP-127b found  a cloud-free atmosphere for this puffy planet with a strong Rayleigh-scattering signature and features compatible with TiO and VO \citep{Palle2017}. The tentative signature of sodium is also claimed and was later confirmed at a $5\sigma$ level by \cite{Chen2018}. \cite{Allart2020} also found signatures of sodium that are compatible both with \cite{Chen2018} and \cite{Spake2021}, but the water feature amplitude analysis at $1.3 \mu$m seems to indicate, at least partially, the presence of clouds.  This is compatible with the upper limits established by \cite{Seidel2020} using simultaneous photometry, but with the less precise HARPS data.
The two observations of WASP-127b with ESPRESSO have both in-transit and out-of-transit data, which allows for a better determination of the activity-induced slope. We chose then to fix the RV semi-amplitude and fit $V_{\rm sys}$, $R_{\rm p}/R_{\rm \star}$, $\xi$, and $\log(\sigma_{\rm W})$, with the priors from Table \ref{table:wasp127priors_chrom}.
\begin{table}
\centering
\caption{Set of priors for the chromatic fits with \texttt{ARoME} for WASP-127b. }
\label{table:wasp127priors_chrom}
\begin{tabular}{lll}
Parameter & Prior \\
\hline
$V_{\rm sys} \,$[km\,s$^{-1}$] & $\mathcal{U}(-10.275,-8.275)$\\
$R_{\rm p} /R_{\rm \star} $ & $\mathcal{G}(0.10103,0.010103)$\\
$\xi \,$ [km] & $\mathcal{G}(0, 1)$\\
$\sigma_{\rm W} \,$ [m\,s$^{-1}$] & $\mathcal{G}(0, 250)$\\
\hline
\hline
\end{tabular}
\end{table}

We decided not to use GPs with this data set, as there are no clear structures seen in the residuals (Fig. \ref{fig:waspnights2}).
The analysis of the transmission spectrum obtained with \texttt{CaRM} (Fig. \ref{fig:wasptransmission}, Table \ref{table:wasp127_radius_table}) does not reveal broadband features, in contrast to what has been found for the previous target. At low resolution, it is compatible with the conclusions of \cite{Palle2017}, \cite{Chen2018}, and \cite{Spake2021}. In fact, the transmission spectrum presented in \cite{Palle2017} has a stronger slope in bluer regions that we cannot probe because it is outside the wavelength range of ESPRESSO. Given the flatness of the retrieved spectrum and large error bars, we did not try to model it with \texttt{PLATON}.

\section{Conclusions}
\texttt{CaRM} is a semi-automatic code that uses the chromatic Rossiter-McLaughlin effect to retrieve the transmission spectra of exoplanets. In this paper, we described the code and studied the  variables to be considered to use this method. \par
We used the code on  HD 189733b with HARPS data and found similar results when compared with previous authors using the same method. The white-light RM fit showed that the model is not fully capable of reproducing the observations. The residuals obtained after the subtraction of the best-fit model show an RM-like pattern that has been previously attributed to spot-crossing events, differential stellar rotation, or CB \citep[][]{Triaud2009, DiGLoria2015, Cegla2016, Dravins2021}. \par
With \texttt{PLATON}, the planetary transmission spectrum was analyzed, and we found that a model containing Rayleigh scattering is preferred over a flat one. We do not have enough evidence with this method to confirm the presence of the haze. Yet, \cite{Sing2015} were able to confirm it using Hubble and Spitzer Space Telescopes observations. \par
For WASP-127b, the white-light fit shows a rather stable system in terms of the systemic velocity. The spin-orbit misalignment found with the classical RM analysis shows a difference  of $\sim 20^\circ$  with the reloaded RM solution. The transmission spectrum, at the resolution \texttt{CaRM} is able to retrieve, is rather unremarkable and with no identifiable features. Despite this, when compared with other literature sources, it is more consistent with a flatter profile \citep{Spake2021} than with a strong Rayleigh scattering in the bluest part of the spectrum \citep[][]{Palle2017,Chen2018}.\par
In this work, we show that \texttt{CaRM} is capable of reproducing both past analysis (with HD189733b) as well as performing new ones with ESPRESSO. However, it is a less sensitive method when compared to broadband transmission spectroscopy from space. The precision on the transmission spectrum depends, in a first degree, on the ratio between the Rossiter-McLaughlin amplitude and the uncertainties on the RV measurements. The uncertainties can be improved by incorporating multiple transits, and better RM models are fundamental to disentangle between the planet and stellar signatures. Taking this in consideration, \texttt{CaRM} proves to be a valuable tool to retrieve and study the transmission spectra of exoplanets using state-of-the-art spectrographs such as ESPRESSO and future HIRES@ELT.\par

\begin{acknowledgements}
The authors acknowledge the ESPRESSO project team for its effort and dedication in building the ESPRESSO instrument. 
This work was supported by FCT - Funda\c{c}\~ao para a Ci\^encia e a Tecnologia through national funds and by FEDER through COMPETE2020 - Programa Operacional Competitividade e Internacionaliza\c{c}\~ao by these grants: UID/FIS/04434/2019; UIDB/04434/2020; UIDP/04434/2020; PTDC/FIS-AST/32113/2017 \& POCI-01-0145-FEDER-032113; PTDC/FIS-AST/28953/2017 \& POCI-01-0145-FEDER-028953; PTDC/FIS-AST/28987/2017 \& POCI-01-0145-FEDER-028987. O.D.S.D. is supported in the form of work contract (DL 57/2016/CP1364/CT0004) funded by national funds through Fundação para a Ciência e Tecnologia (FCT). The INAF authors acknowledge financial support of the Italian Ministry of Education, University, and Research with PRIN 201278X4FL and the "Progetti Premiali" funding scheme). N.J.N acknowledges, additionally, support from FCT through the project CERN/FIS-PAR/0037/2019. J.H.C.M. is supported in the form of a work contract funded by Fundação para a Ciência e Tecnologia (FCT) with the reference DL 57/2016/CP1364/CT0007; and also supported from FCT through national funds and by FEDER-Fundo Europeu de Desenvolvimento Regional through COMPETE2020-Programa Operacional Competitividade e Internacionalização for these grants UIDB/04434/2020 \& UIDP/04434/2020, PTDC/FIS-AST/32113/2017 \& POCI-01-0145-FEDER-032113, PTDC/FIS-AST/28953/2017 \& POCI-01-0145-FEDER-028953, PTDC/FIS-AST/29942/2017.
R. A. is a Trottier Postdoctoral Fellow and acknowledges support from the Trottier Family Foundation. This work was supported in part through a grant from FRQNT. This work has been carried out within the framework of the National Centre of Competence in Research PlanetS supported by the Swiss National Science Foundation. The authors acknowledge the financial support of the SNSF. 
ASM, JIGH, CAP and RR acknowledge financial support from the Spanish Ministry of Science and Innovation (MICINN)  project PID2020-117493GB-I00, and from the Government of the Canary Islands project ProID2020010129.
JIGH also acknowledges financial support from the Spanish MICINN under 2013 Ram\'on y Cajal program RYC-2013-14875.
This work has been carried out in the frame of the National Centre for Competence in Research ``PlanetS'' supported by the Swiss National Science Foundation (SNSF). This project has received funding from the European Research Council (ERC) under the European Union's Horizon 2020 research and innovation programme (project {\sc Spice Dune}, grant agreement No 947634, project {\sc SCORE}, grant agreement No 851555).
V.A. acknowledges the support from FCT through Investigador FCT contract nr. IF/00650/2015/CP1273/CT0001.

\end{acknowledgements}
 
\bibliographystyle{aa}
\bibliography{crm}
\begin{appendix}\label{appendix}

\section{The limb-darkening laws}\label{ldlaws}
There are several parametrizations of limb darkening which can be more adequate depending on the physical properties of the star. Some of those parametrizations, which can be used in \texttt{CaRM}, are the linear (eq. \ref{eq:eq1}), the quadratic (eq. \ref{eq:eq2}), and the nonlinear (eq. \ref{eq:eq3}):

\begin{center}
\begin{equation}\label{eq:eq1}
    \frac{I(\mu)}{I(\mu=1)} = 1-u_1 (1-\mu);
\end{equation}    
\end{center}

\begin{center}
\begin{equation}\label{eq:eq2}
    \frac{I(\mu)}{I(\mu=1)} = 1-u_1 (1-\mu) - u_2 (1-\mu)^2;
\end{equation}
\end{center}

\begin{figure}[!ht]
  \centering
  \includegraphics[width=\linewidth]{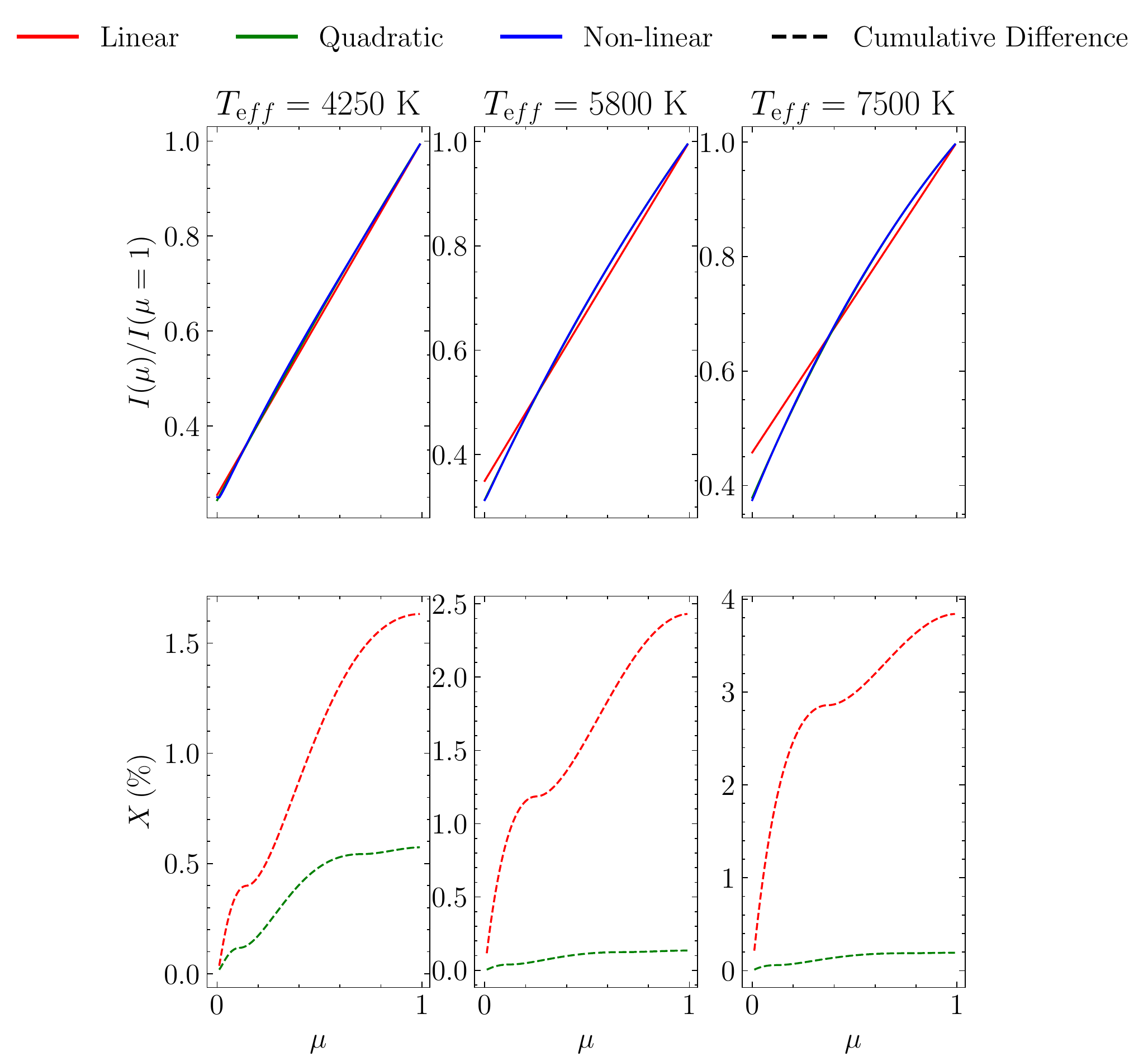}
  \caption{Stellar temperature impact in the limb-darkening laws. Top row: Comparison between the normalized intensity for a linear and quadratic, and nonlinear limb-darkening laws (top). Bottom row: Cumulative percentage difference for the linear and quadratic laws and the nonlinear. }
  \label{fig1:profdif}
\end{figure}

\begin{center}
    \begin{equation}\label{eq:eq3}
    \frac{I(\mu)}{I(\mu=1)} = 1-\sum_{n=1}^{4} u_n (1-\mu^{n/2}).
\end{equation}
\end{center}
The normalized intensities show the largest difference for $\mu = 0$ for the three temperatures, Fig.\ref{fig1:profdif}, with $\mu$ representing the cosine of the angle between the observer line of view and the normal to the stellar surface. We used the cumulative function of the difference between models to show, not only the localized differences expressed by larger slopes, but also the weight they have on the total difference. \par
For the mock system we used, the cumulative difference between the quadratic and nonlinear laws is no larger than $4 \%$, and it increases with increasing stellar temperature. This is similar to the trend that is observed in RVs and shows that a correct modeling of the limb-darkening law is important for future instruments with higher RV precision.\par

\section{Simulated system parameters}
\begin{table}[H]
\centering
\caption{Simulated planetary system properties for a hot-Jupiter-like planet. The variables that serve as input to \texttt{LDTk} have typical uncertainties associated with them.}
\begin{tabular}{ll}
\hline
Variable & Value \\
\hline
$R_\star \,[R_{\rm \odot}]$ & $1.0$ \\

$T_{\rm eff} \,$[K] & $(4250,\,5800,\,7500) \pm 50$ \\

$\log(g)$ & $4.44 \pm 0.04$\\

[Fe/H] [dex] & $0.0 \pm 0.02$\\

$P \,$[days] & $4.0$ \\

$a \,[R_\star]$ & $4.0$\\

$\lambda \,[^\circ]$ & $0,\,30,\,60$ \\

$v\,\sin \,i_\star \,$[km\,s$^{-1}$] & $1.7$  \\

$R_{\rm p}/R_{\rm *}$ & $0.10$ \\

$i \,[^\circ]$ & $ 85.72$ \\

$\sigma_0 \,$[km\,s$^{-1}$] & $ 4.06$ \\

$\beta_0 \,$[km\,s$^{-1}$] & $3.7$ \\

$\zeta_t \,$[km\,s$^{-1}$] & $0$ \\
\hline
\end{tabular}
\label{table:tab1}
\end{table}

\section{Transmission spectra tables}
\begin{table}[H]
    \centering
    \caption{Planet radius, in stellar radius units, obtained by our analysis as a function of wavelength ($\lambda_c$ and bin width $\Delta \lambda$) for HD 189733b. The uncertainties in the radius correspond to a $68 \%$ confidence interval. For the analysis of the transmission spectrum, we subtracted 0.0277 $R_{\rm p}/R_{\star}$ to match the average radius ratio of \cite{Pont2013}.}
    \label{table:hd189_radius_table}
    \begin{tabular}{llll}
        $\lambda_c$ & $\Delta \lambda$ & $R_{\rm p} / R_{\rm \star}$ & Uncertainty \\
        \hline
        399.71&21.75&0.1871&{+0.0015}/{-0.0016}\\
        444.68&25.03&0.1850&$\pm{0.0011}$\\
        494.75&26.69&0.1840&$\pm{0.0009}$\\
        544.88&24.83&0.18178&{+0.0010}/{-0.0009}\\
        598.23&29.60&0.1832&$\pm{0.0011}$\\
        658.695&24.925&0.1845&$\pm{0.0018}$\\
        \hline
        \hline
    \end{tabular}
\end{table}

\begin{table}[!htbp]
    \centering
    \caption{Same as Table \ref{table:hd189_radius_table}, but for WASP-127b. No shift was applied.}
    \label{table:wasp127_radius_table}
    \begin{tabular}{llll}
       $\lambda_c$ & $\Delta \lambda$ & $R_{\rm p} / R_{\rm \star}$ & Uncertainty \\
        \hline
        398.345 & 21.195 &0.0950&${+0.0060}/{-0.0062}$\\
        443.335 & 27.425 &0.0982&${+0.0052}/{-0.0053}$\\
        492.665 & 25.515 &0.0995&${+0.0051}/{-0.0050}$\\
        530.270 & 15.550 &0.1074&${+0.0058}/{-0.0059}$\\
        567.730 & 25.320 &0.0994&${+0.0063}/{-0.0062}$\\
        622.410 & 26.81 &0.1044&${+0.0060}/{-0.0059}$\\
        673.115 & 19.855 &0.1016&${+0.0069}/{-0.0072}$\\
        761.325 & 29.315 &0.1003&${+0.0066}/{-0.0071}$\\
        \hline
        \hline
    \end{tabular}
\end{table}

\section{Individual nights fits}
\begin{figure*}
\centering

\includegraphics[width=.49\linewidth]{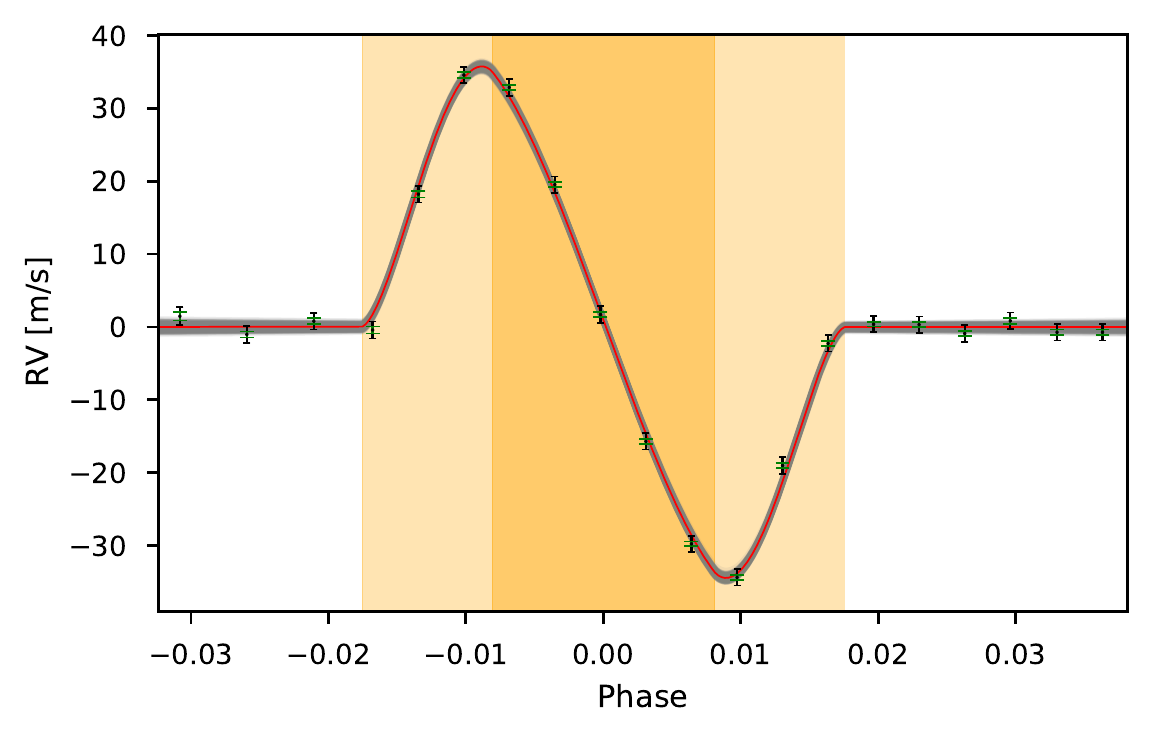}
\includegraphics[width=.49\linewidth]{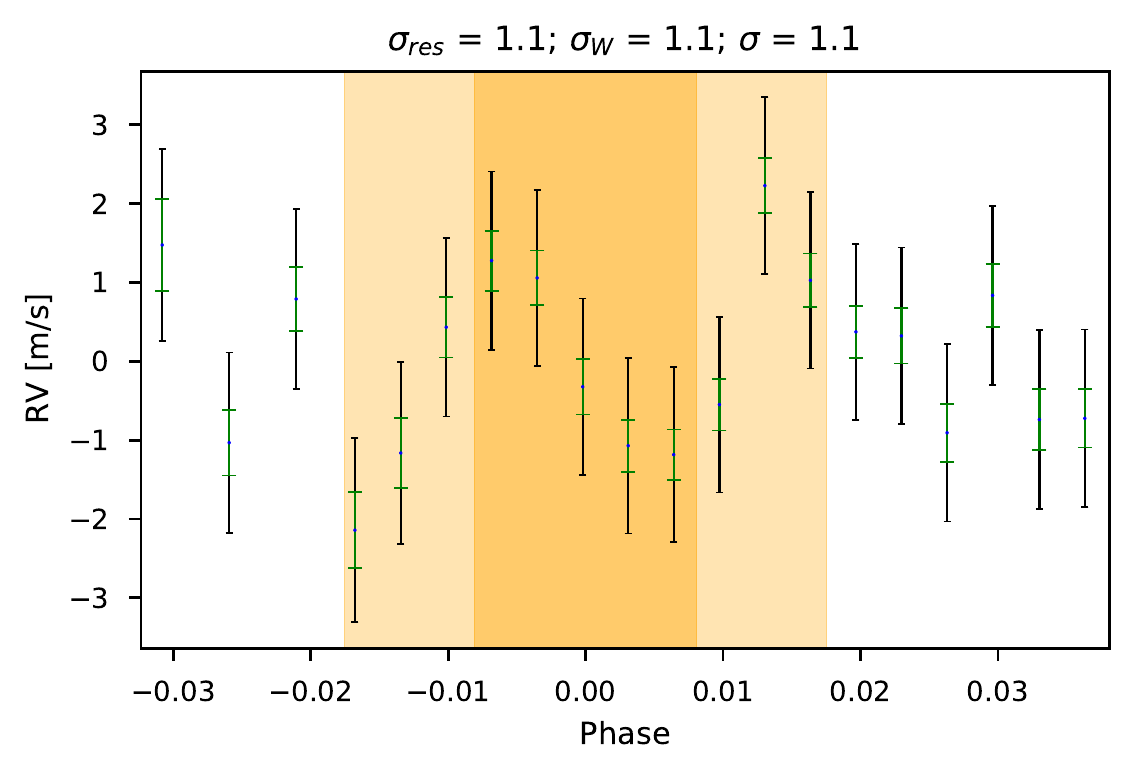}
\includegraphics[width=.49\linewidth]{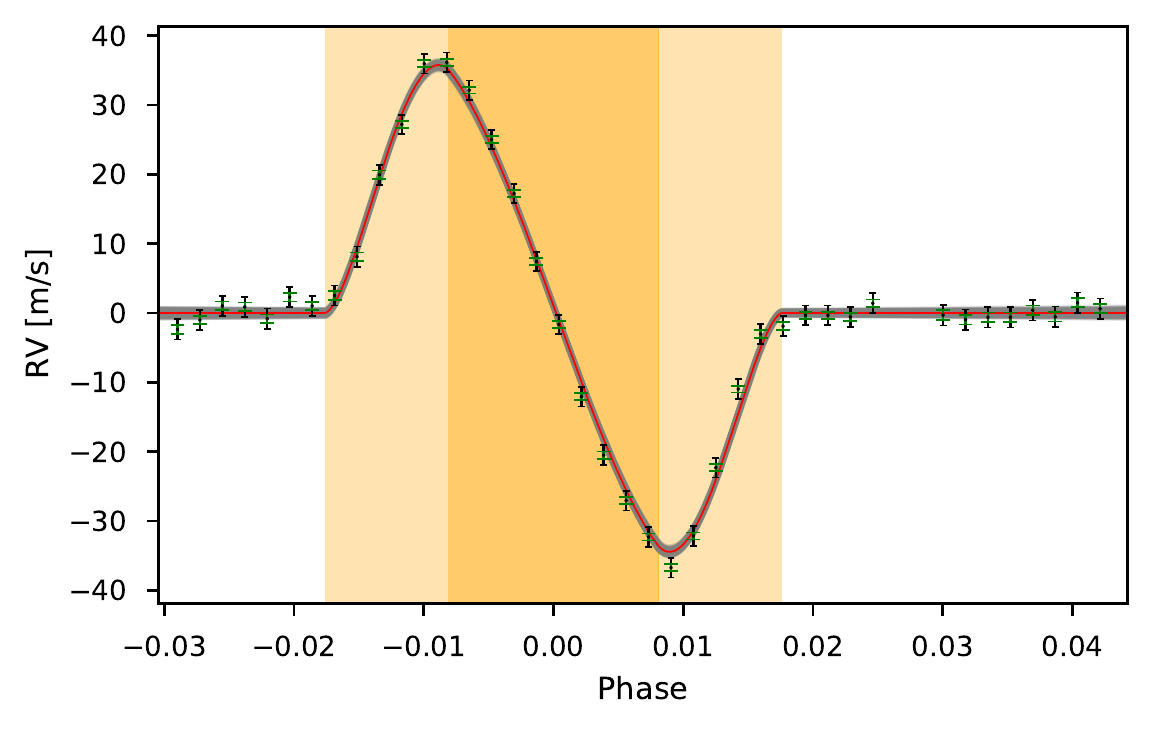}
\includegraphics[width=.49\linewidth]{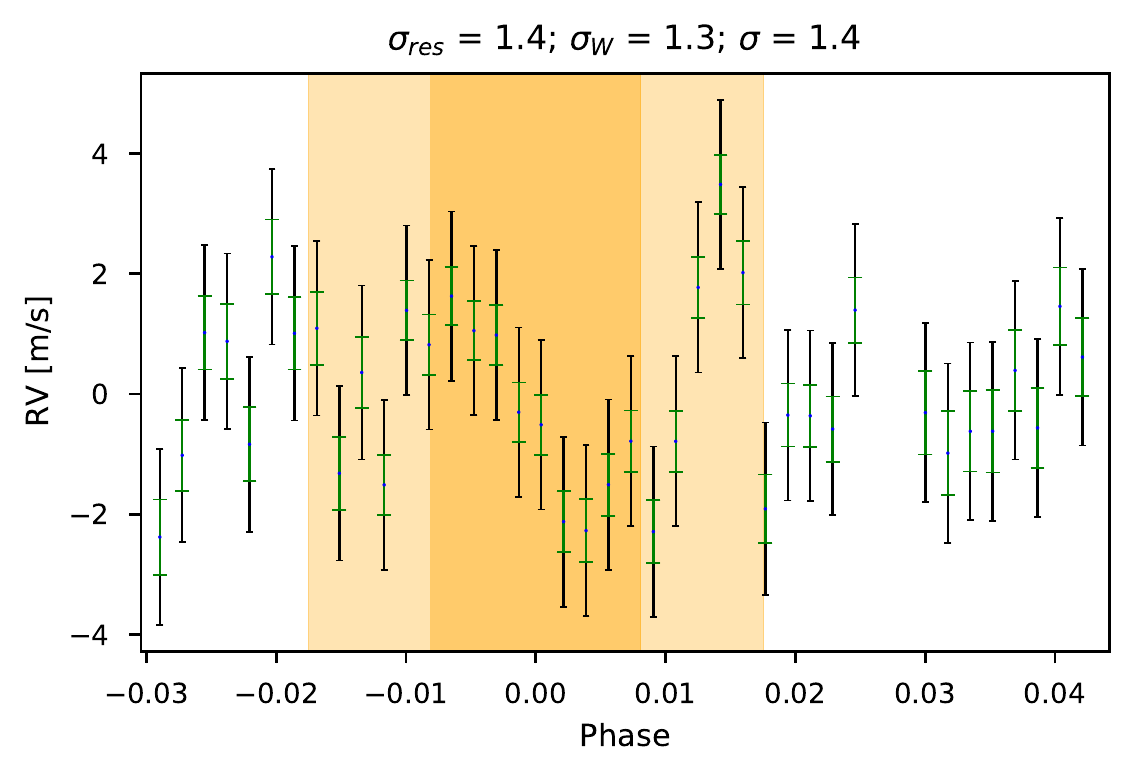}
\includegraphics[width=.49\linewidth]{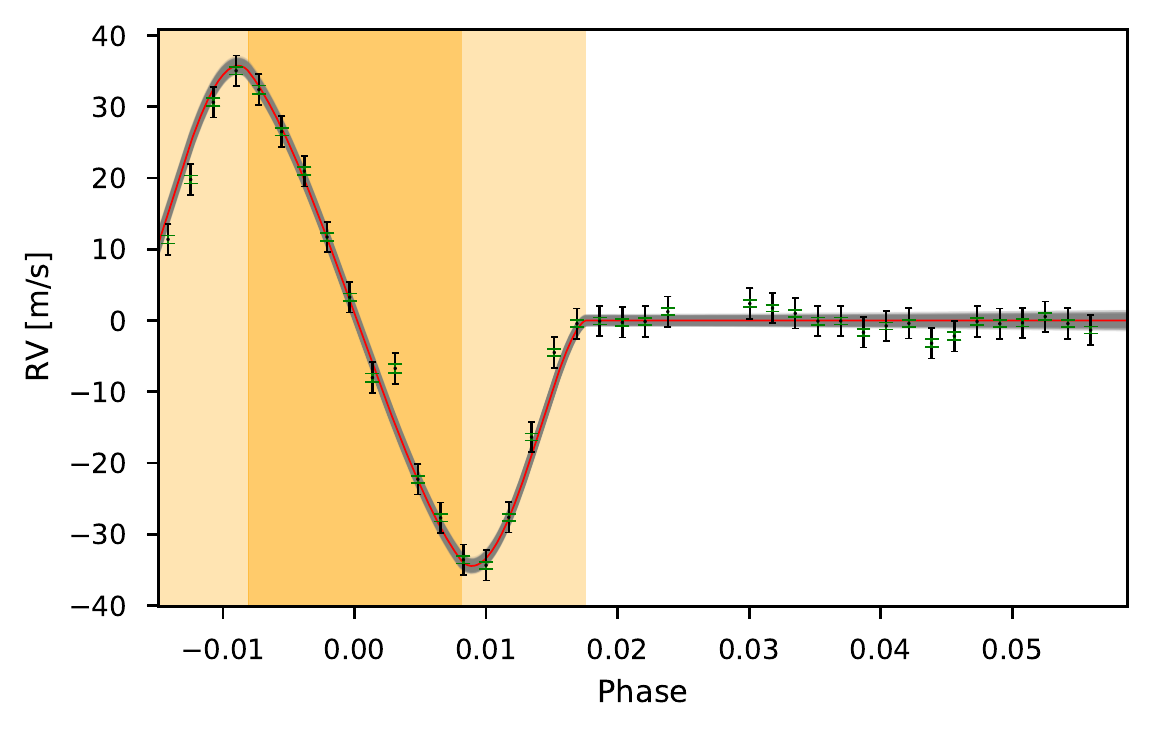}
\includegraphics[width=.49\linewidth]{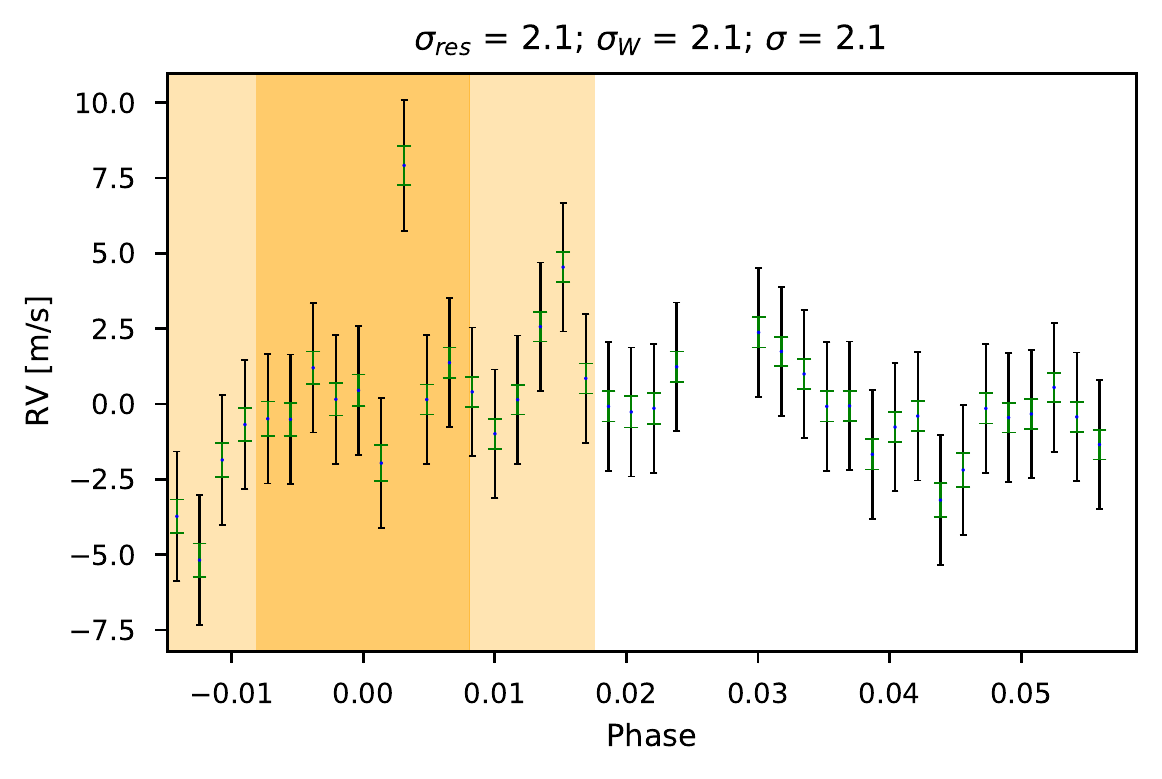}

\caption{Data fit and residuals. The elements in the plots have the same meaning as in Fig. \ref{fig:hd189jres}.}

\label{fig:hd189nights}

\end{figure*}

\begin{figure*}[ht!]
\centering
  \includegraphics[width=.49\linewidth]{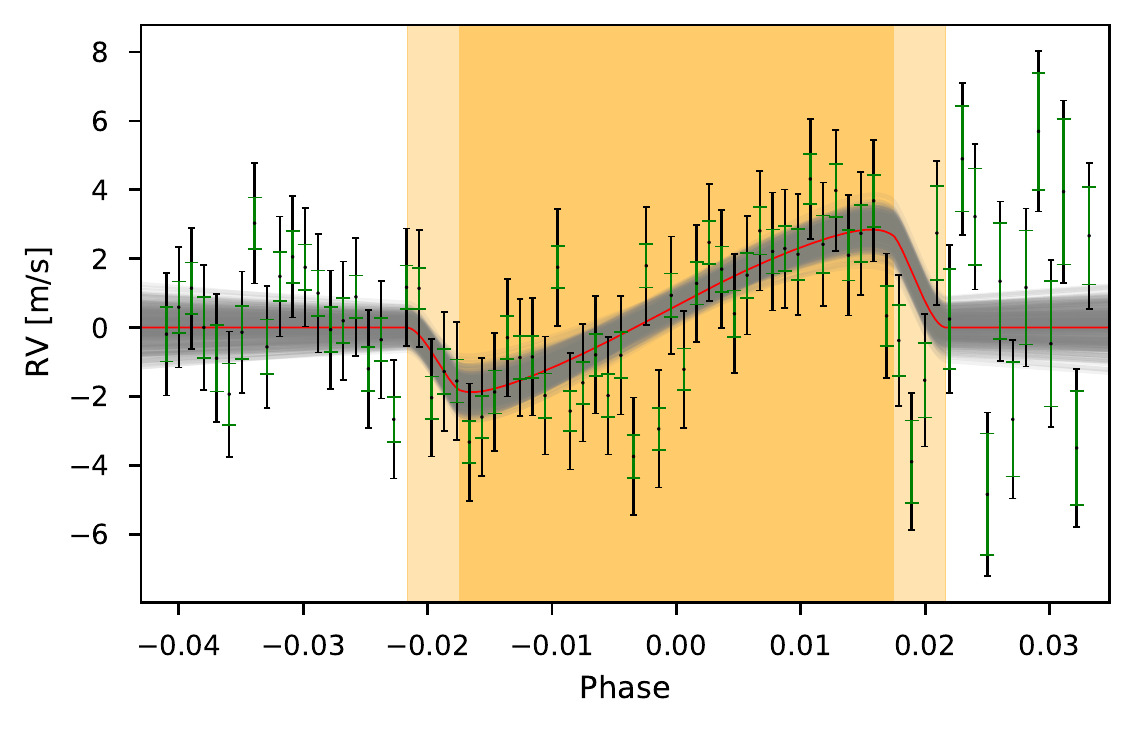}
  \includegraphics[width=0.49\linewidth]{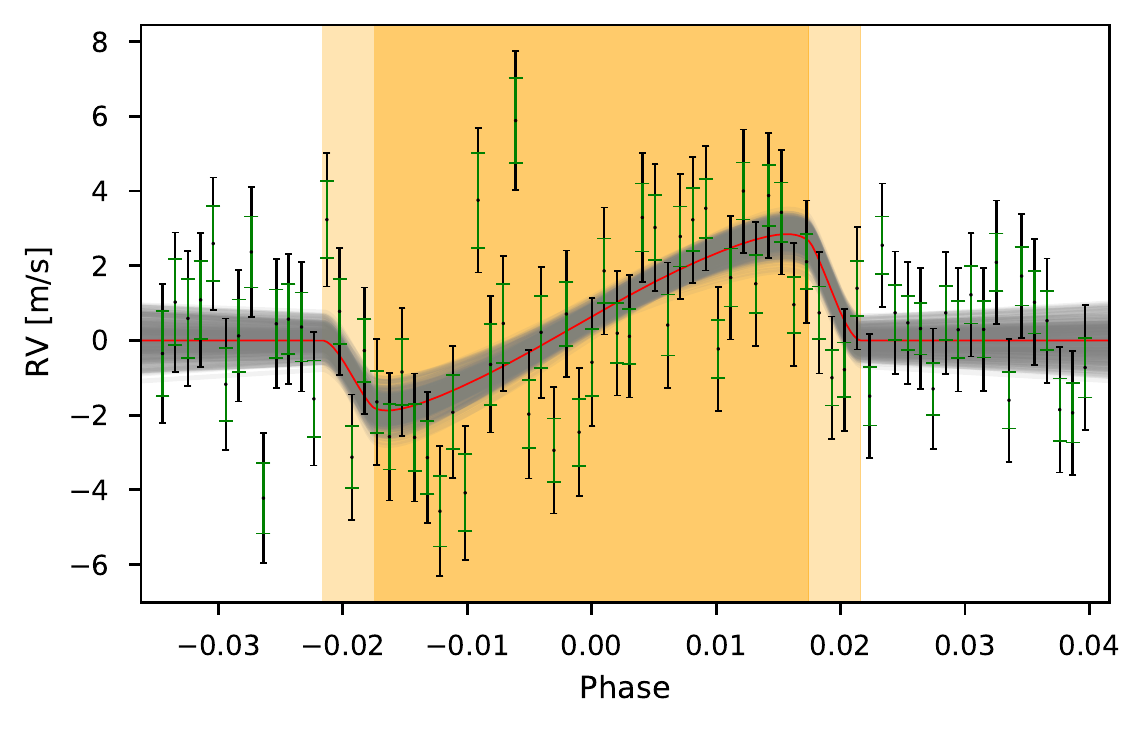}
  \includegraphics[width=.49\linewidth]{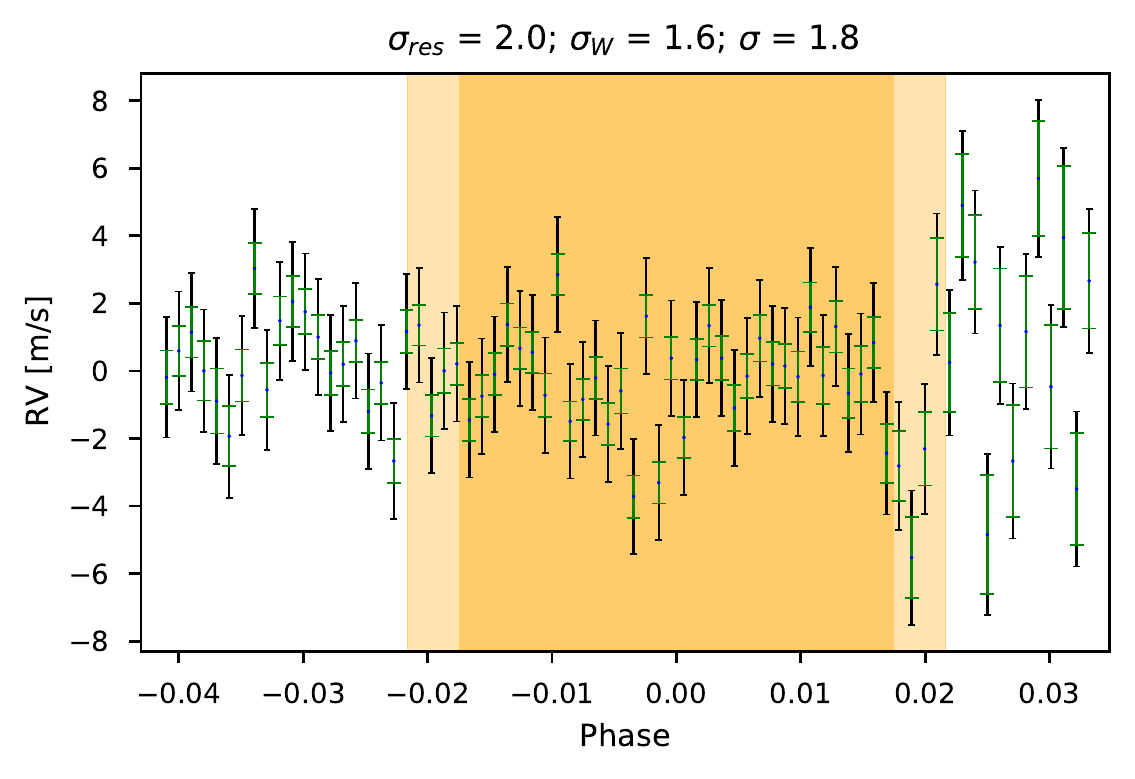}
  \includegraphics[width=0.49\linewidth]{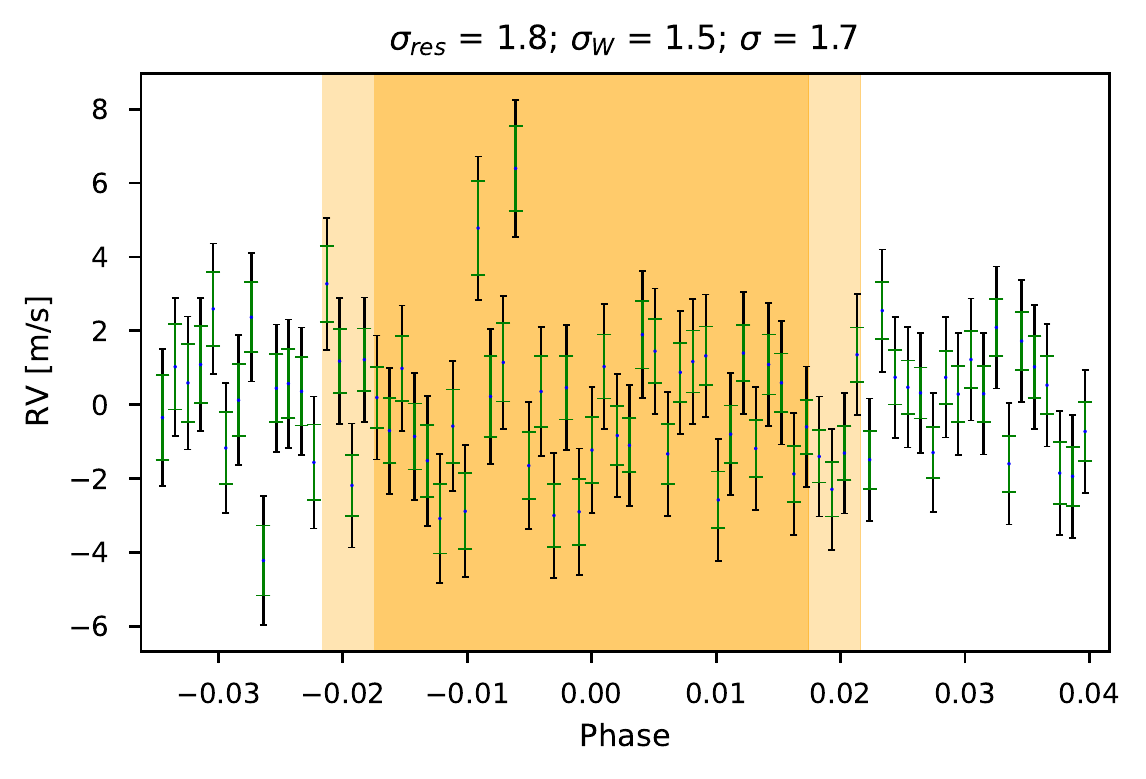}
  \caption{Data fit and residuals. The elements in the plots have the same meaning as in Fig. \ref{fig:hd189jres}.}
\label{fig:waspnights2}
\end{figure*}
\clearpage

\onecolumn
\section{Reference values table}
\begin{table}[h!]
\centering
\caption{Reference values for the retrievals, \texttt{CaRM} and \texttt{PLATON}. }
\label{table:tablepars}
\begin{tabular}{lllll}
\multirow{2}{*}{} & \multicolumn{2}{l}{\textbf{HD189733}} & \multicolumn{2}{l}{\textbf{WASP-127}} \\
 \hline
Parameter & Value & Source & Value & Source\\
\hline
$R_\star\,[R_{\rm \odot}]$  & $0.766^{+0.007}_{-0.013}$                 & \cite{Triaud2009} &   $1.303 \pm 0.037$   &\cite{Allart2020} \\
$T_{\rm eff} \,$[K]         & $4969 \pm 43$                                     & \cite{Sousa2018}  &   $5842 \pm 14$       &\cite{Allart2020} \\
$\log(g)$                               & $4.60 \pm 0.01$                                                   & \cite{Sousa2018}   &$4.23 \pm 0.02$        &\cite{Allart2020} \\
$\rm [Fe/H]$ [dex]       & $-0.07 \pm 0.02$                                                 & \cite{Sousa2018}   &$-0.19 \pm 0.01$       &\cite{Allart2020} \\
$T_0 \,$[MBJD]              & $53988.30339 ^{+0.000072}_{-0.000039}$    & \cite{Triaud2009} &$2456776.621238 \pm 0.00023181$ &\cite{Seidel2020}\\
$P \,$[days]                & $2.21857312 ^{+0.00000036}_{-0.00000076}$ & \cite{Triaud2009} & $4.17806203 \pm 0.00000088$   &\cite{Seidel2020} \\
$a \,[R_\star]$                 & $8.756 ^{+0.0092}_{-0.0056}$                      & \cite{Triaud2009} &$7.808 \pm 0.109$      &\cite{Seidel2020} \\
$\lambda \,[^\circ]$            & $-0.85^{+0.28}_{-0.32}$                                   & \cite{Triaud2009} &$-128.41^{+5.60}_{-5.46}$ & \cite{Allart2020}\\
$v \sin(i_{\rm \star}) \,$[km\,s$^{-1}$] & $3.05$                                                                                & \cite{Triaud2009}             &$0.53^{+0.07}_{-0.05}$& \cite{Allart2020}\\
$R_{\rm p} \,[R_\star]$                                         & $0.1581 \pm 0.0005$                             & \cite{Triaud2009}             &$0.10103441 \pm 0.00047197$                 &\cite{Seidel2020}\\
$K \,$[m\,s$^{-1}$]                                     & $ 201.96 ^{+1.07}_{-0.63} $               & \cite{Triaud2009}                     & $21.51 \pm 2.78$                                                         &\cite{Seidel2020}\\
$i \,[^\circ]$                                  & $85.508 ^{+0.10}_{-0.05}$                                 & \cite{Triaud2009}                             & $87.85 \pm 0.35$&\cite{Seidel2020}\\
$\sigma_0 \,$[km\,s$^{-1}$]             & $ 3.10 \pm 0.01$                                                                              & This work       \tablefootmark{a}               &$3.146\pm 0.002$ & This work\tablefootmark{a}\\
$\beta_0 \,$[km\,s$^{-1}$]              & $2.518 \pm 0.001$                                                                             & This work \tablefootmark{b}             & $3.313 \pm 0.004$& This work\tablefootmark{b}\\
$\zeta_t \,$[km\,s$^{-1}$]                      & $4.0$                                                                                 & \cite{DiGLoria2015}             &$3.79 \pm 0.05$ & This work \tablefootmark{c}\\
\hline
\end{tabular}
\tablefoot{\tablefoottext{a}{Computed using the average standard deviation of the Gaussian fit to the out-of-transit measurements.}
\tablefoottext{b}{Estimated as described in \citet{2002A&A...392..215S}. }
\tablefoottext{c}{Computed with the calibration of \cite{Doyle2014} and valid for stars with $T_{\rm eff}$ from $5200K$ to $6400K$ and $\log(g)$ between $4.0$ and $4.6$.}}
\end{table}

\clearpage
\onecolumn
\section{Corner plot of HD 189733b}
\begin{figure*}[!htb]
  \centering
  \includegraphics[width=0.95\linewidth]{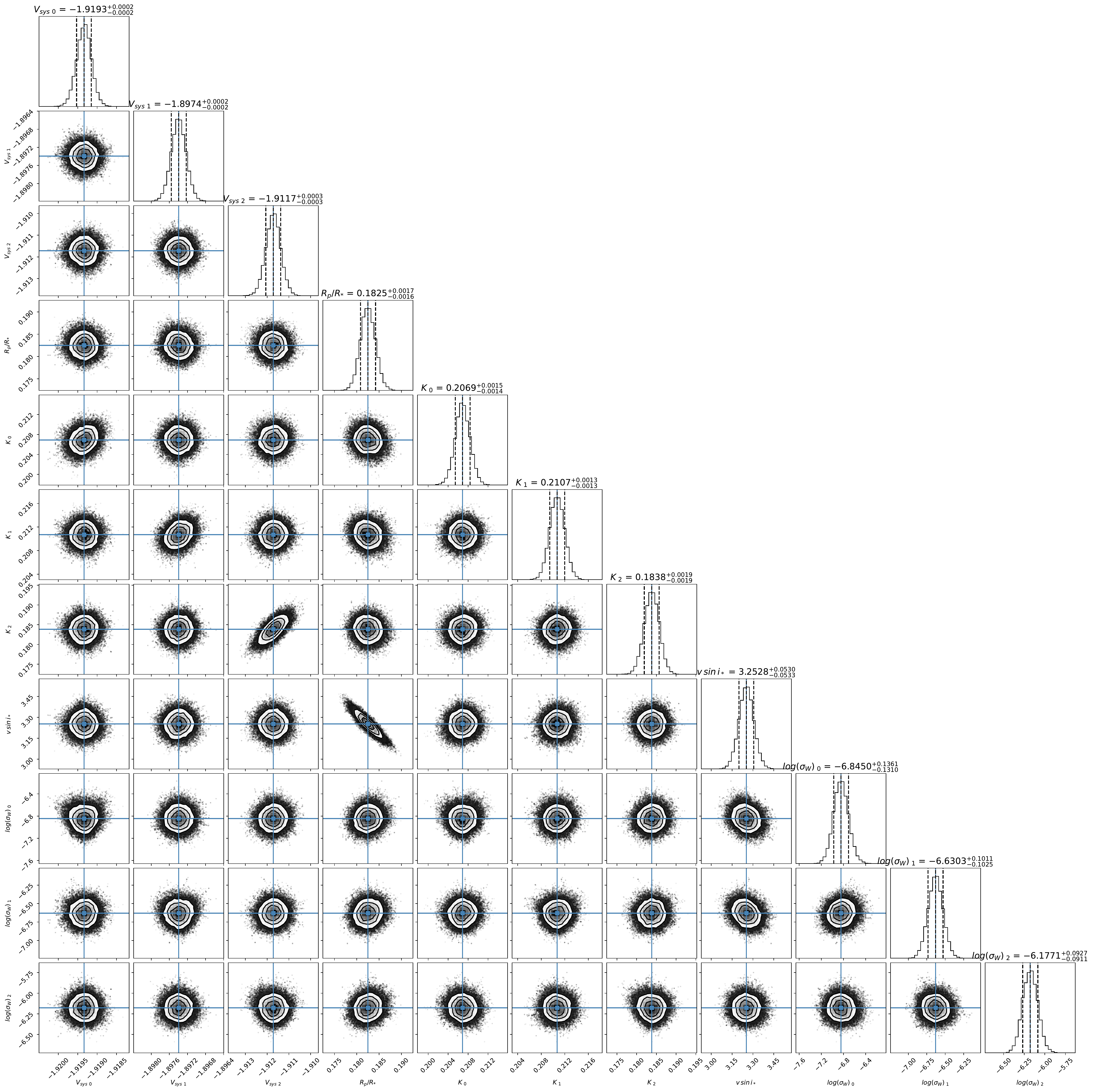}
  \caption{Corner plot of the joint white-light fit for the three HARPS observations of HD 189733b. The values and error bars represent the median and $1\sigma$ uncertainty for the posterior distributions. The number under the identification of each variable, when present, represents an individual fit of that particular parameter.}
  \label{fig:hdcorner}
\end{figure*}
\clearpage

\section{Corner plot of WASP-127b}

\begin{figure*}[!htb]
  \centering
  \includegraphics[width=0.95\linewidth]{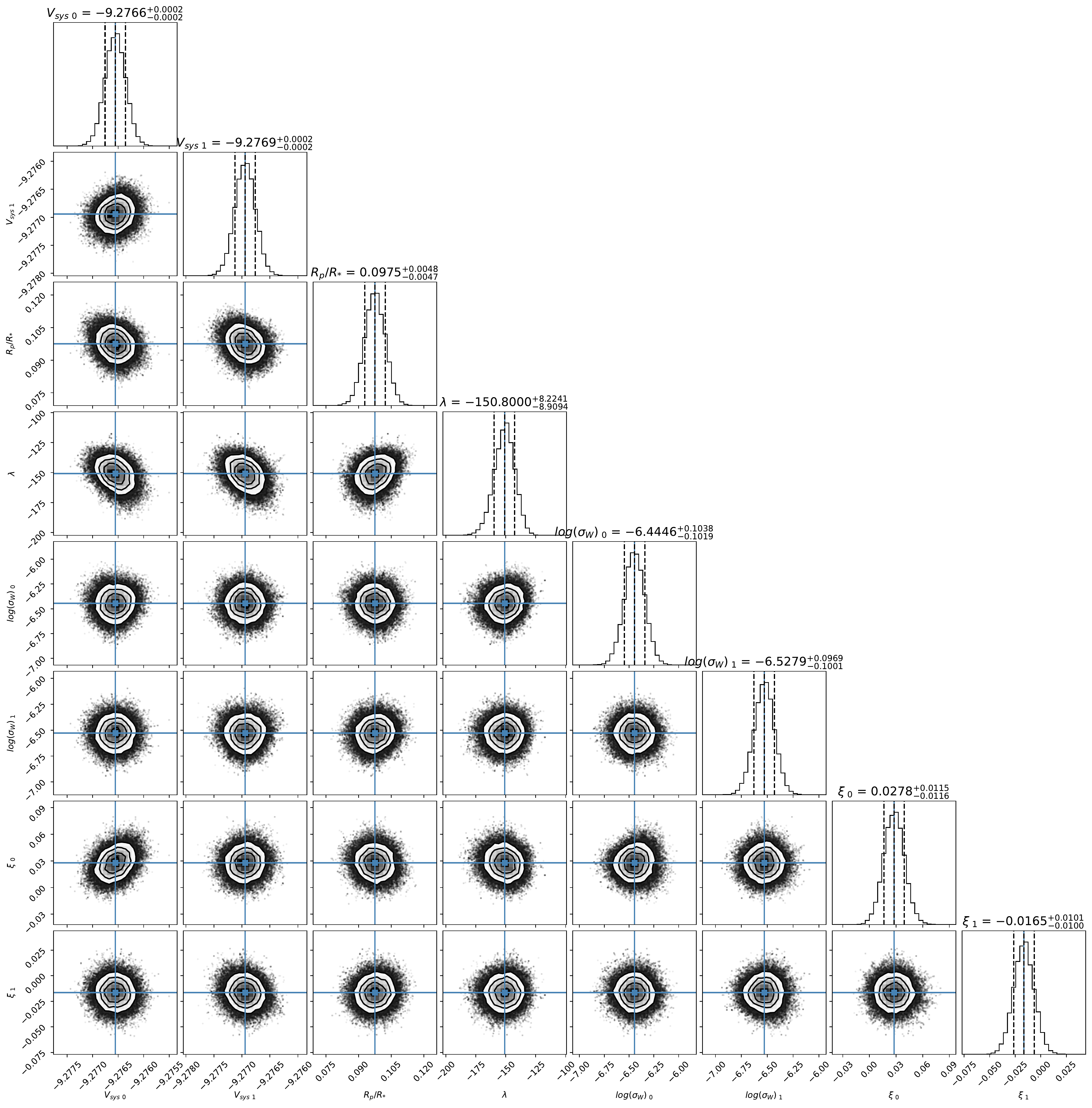}
  \caption{Corner plot of the joint white-light fit for the two ESPRESSO observations of WASP-127b.}
  \label{fig:waspcorner}
\end{figure*}
\end{appendix}
\end{document}